\documentclass[apj]{emulateapj}
\usepackage{amsmath}

\def\mpch {h$^{-1}$Mpc} 
\def\hmsun{\hbox{h$^{-1}M_{\odot}$}}
\def\msun{\hbox{$M_{\odot}$}}
\def \ergs{erg s$^{-1}$}

\shorttitle{HOD of X-ray AGN}
\shortauthors{Allevato et al.}
\begin{document}

\slugcomment{Accepted for publication in The Astrophysical Journal}

\title{Occupation of X-ray selected galaxy groups by X-ray AGN}

\author{V. Allevato\altaffilmark{1,2,3}, 
A. Finoguenov\altaffilmark{4,5},  
G. Hasinger\altaffilmark{6}, 
T. Miyaji\altaffilmark{7,8}, 
N. Cappelluti\altaffilmark{9,6}, 
M. Salvato\altaffilmark{2,3},
G. Zamorani\altaffilmark{9}, 
R. Gilli\altaffilmark{9}, 
M. R. George\altaffilmark{12,13}, 
M. Tanaka\altaffilmark{10},
%A. Leauthaud\altaffilmark{9,11},
M. Brusa\altaffilmark{2}, 
%A. Bongiorno\altaffilmark{2}, 
%J. B. James\altaffilmark{10,11}, 
%H. J. McCracken\altaffilmark{9}, 
 J. Silverman\altaffilmark{10},
 F. Civano\altaffilmark{14},
 M. Elvis\altaffilmark{14} 
 F. Shankar\altaffilmark{15} }
%O. Le Fevre\altaffilmark{15}, 
%J.P. Kneib\altaffilmark{15}, 
%L. Guzzo\altaffilmark{16}, 

%S. Mei\altaffilmark{17,13}, 
%K. Bundy\altaffilmark{18} and 

\altaffiltext{1}{Max-Planck-Institut f$\ddot{u}$r Plasmaphysik, Boltzmannstrasse 2, D-85748 Garching, Germany}
\altaffiltext{2}{Max-Planck-Institute f$\ddot{u}$r Extraterrestrische Physik, Giessenbachstrasse 1, D-85748 Garching, Germany}
\altaffiltext{3}{Excellence Cluster Universe, Boltzmannstrasse 2, D-85748 Garching, Germany}
\altaffiltext{4}{Department of Physics, University of Helsinki, Gustaf H\"allstr\"omin katu 2a, FI-00014 Helsinki, Finland}
\altaffiltext{5}{University of Maryland, Baltimore County, 1000 Hilltop Circle, Baltimore, MD 21250, USA}
\altaffiltext{6}{Institute for Astronomy, University of Hawaii, 2680 Wood- lawn Drive, Honolulu, HI, 96822 USA.}
%\altaffiltext{4}{California Institute of Technology, 1201 East California Boulevard, Pasadena, 91125, CA}
\altaffiltext{7}{Instituto de Astronomia, Universidad Nacional Autonoma de Mexico, Ensenada, Mexico (mailing adress: PO Box 439027, San Ysidro, CA, 92143-9024, USA)}
\altaffiltext{8}{Center for Astrophysics and Space Sciences, University of California at San Diego, Code 0424, 9500 Gilman Drive, La Jolla, CA 92093, USA}
\altaffiltext{9}{INAF-Osservatorio Astronomico di Bologna, Via Ranzani 1, 40127 Bologna, Italy}
%\altaffiltext{10}{Astronomy Department, University of California, Berkeley, 601 Campbell Hall, Berkeley CA, 94720-7450, USA}
%\altaffiltext{11}{Dark Cosmology Centre, University of Copenhagen, Juliane Maries Vej 30, 2100 Copenhagen, Denmark}
\altaffiltext{10}{Institute for the Physics and Mathematics of the Universe, The University of Tokyo, 5-1-5 Kashiwanoha, Kashiwa-shi, Chiba 277-8583, Japan}
%\altaffiltext{9}{Max-Planck-Institute f$\ddot{u}$r Astrophysik, Karl-Schwarzschild-Str., D-85748 Garching, Germany}
%\altaffiltext{15}{Laboratoire d’Astrophysique de Marseille, CNRS-Univ. de Provence, Technopole de Marseille-Etoile, France}
%\altaffiltext{16}{INAF-Osservatorio Astronomico di Brera, Merate/Milano, Italy}
%\altaffiltext{17}{University of Paris Denis Diderot, 75205 Paris Cedex 13, France-GEPI}
%\altaffiltext{13}{Observatoire de Paris, Section de Meudon, 5 Place J. Janssen, 92195 Meudon Cedex, France}
%\altaffiltext{18}{Astronomy Department, University of California, Berkeley, CA 94705, USA}
\altaffiltext{11}{Lawrence Berkeley National Laboratory, 1 Cyclotron Road, Berkeley CA 94720, USA}
\altaffiltext{12}{Berkeley Center for Cosmological Physics, University of California, Berkeley, CA 94720, USA}
\altaffiltext{13}{Department of Astronomy, University of California, Berkeley, CA 94720, USA}
\altaffiltext{14}{Smithsonian Astrophysical Observatory, 60 Garden St., Cambridge, MA 02138, USA}
\altaffiltext{15}{GEPI, Observatoire de Paris, CNRS, Univ. Paris Diderot, 5 Place Jules Janssen, 92195 Meudon, France}
\begin{abstract}

We present the first direct measurement of the mean Halo Occupation Distribution (HOD) of X-ray
  selected AGN in the COSMOS field at $z \leq 1$, based on the association of 41 XMM
  and 17 C-COSMOS AGN with member galaxies of 189 X-ray detected galaxy groups 
  from XMM and \textit{Chandra} data.%  We identify 41 XMM-COSMOS AGN members of galaxy groups defined 
%  as AGN located within 3 times the group line-of-sight velocity dispersion 
%  and within $R_{200}$ and 17 additional sources including in the 
%  analysis C-COSMOS only selected AGN.
  %We develop a method to properly correct the AGN HOD 
  %for the redshift evolution of the AGN 
  %number density in each halo mass bin and 
  %for the fact that we are working with an AGN flux-limited sample. 
  We model the mean AGN occupation in the halo mass range 
  $logM_{200}[M_{\odot}] = 13- 14.5$ 
  with a rolling-off power-law with the best fit index
  %Fitting the AGN HOD with a power-law model in the halo mass range
  %$logM_{200}[M_{\odot}] = 13- 14.5$, we constrain 
  $\alpha=0.06(-0.22; 0.36)$ 	
  and normalization parameter $f_a=0.05(0.04; 0.06)$.
  %at $z=0$ and
  %$log \langle L_{X,AGN} \rangle = 42.3$ erg/s.
  We find the mean HOD of AGN among central galaxies to be modelled
  by a softened step function at log$M_h>$log$M_{min}=12.75(12.10, 12.95)$\msun\
  while for the satellite AGN HOD 
  we find a preference for an increasing AGN fraction with $M_h$ 
  suggesting that the average number of AGN in satellite galaxies grows slower ($\alpha_s < 0.6$) than the 
  linear proportion ($\alpha_s=1$) observed for the satellite
  HOD of samples of galaxies.
  We present an estimate of the projected auto correlation function (ACF)
  of galaxy groups over the range of $r_p =$ 0.1 - 40 \mpch\
  at $\langle z \rangle=0.5$.
We use the large-scale clustering signal
  to verify the agreement between
  the group bias estimated by using the observed galaxy groups
  ACF and the value derived from the group mass estimates.
  We perform a measurement of the projected AGN-galaxy group cross-correlation function,
  excluding from the analysis AGN that are within galaxy groups and we model the 2-halo term
  of the clustering signal with the mean AGN HOD based on our results.

\end{abstract}


\section{Introduction}\label{sec:intro}

In the past decade it has become clear that the majority of galaxies have likely 
had one or more brief active periods, shifting our view of astrophysical 
black holes (BHs) from the role of exotic phenomena to 
fundamental ingredients of cosmic structures. The question of when and under 
which physical conditions the active galactic nuclei (AGN) activate is important for understanding 
not only the origin and evolution of BHs but also the origin and 
evolution of galaxies.

The clustering analysis can powerfully test theoretical model predictions 
and address which physical processes trigger AGN activity. In the 
framework of the cold dark matter structure formation (CDM), the spatial distribution 
of AGN in the Universe (described by the correlation function, CF) can be related to how AGN are 
biased with respect to the underlying matter distribution and to the typical mass of 
dark matter halos (DMHs) in which they reside. 
There have been several studies of the bias evolution of optical quasars with 
redshift, based on survey such as 2QZ and SDSS
\citep{Cro05,Por06,Shen09, Ros09}. All the previous studies infer the picture 
that the quasar bias evolves with redshift 
following a constant mass evolution in 
the range log$M_{h}$[\hmsun] $\sim 12.5-13$, i.e. halo masses similar to group scales, 
where the combination of low velocity dispersion and moderate galaxy space density 
yields to the highest probability of a close encounter \citep{Hop08, McI09}.
Models of major mergers between gas-rich galaxies appear to naturally produce
many observed properties of quasars, such as
the shape and the evolution of the quasar luminosity function and the 
large-scale quasar clustering as a function of luminosity and redshift \citep{Hop07,Hop08, Shen09a, Sha09, Sha10, Bon09},
supporting the scenario in which major mergers dominate the bright quasar populations.
On the other hand, the majority of the results on the clustering of X-ray selected AGN, 
suggest a picture where moderate-luminosity AGN live in massive DMHs 
($12.5< logM_{h}$[\hmsun] $<13.5$) up to $z \sim 2$ 
\citep{Gil05, Yan06, Gil09, Hic09, Coi09, Kru10, Cap10, Kru10, Alle11,Miy11},
i.e. X-ray selected AGN samples appear to cluster more strongly than luminous quasars
\citep[see][for a review on the argument]{Cap12}.
The reason for this is not completely clear, but several studies argued that
this large bias and DMH masses could suggest a different AGN triggering 
mechanism with respect to bright quasars characterized by galaxy merger-induced fueling.
Several studies on the morphology of the AGN host galaxies 
have demonstrated that major mergers of galaxies 
are not likely to be the single dominant mechanism responsible for 
triggering AGN activity at low ($z \sim 1$) \citep{Geo07,Sil09,Geo09,Dun03, Gro05,
Pie07, Gab09, Rei09, Tal09,Cis11,Sil11} and high redshift ($z \sim$ 2) \citep{Ros11,Koc11,Sch11}.

The theoretical understanding of galaxy clustering and bias factor has been greatly 
enhanced through the Halo Occupation Distribution (HOD) framework. 
%\citep{Kau97,Pea00,Coo02,Tin05,Zhe05}.
In this framework, the virialised DMH with typical 
overdensities of $\Delta_{200}$ (defined w.r.t. the mean density)
%expected to be in approximate dynamical equilibrium and 
are described in terms of the probability $P(N|M)$ of a halo of 
given mass $M$ of having $N$ galaxies. 
A simple way to model the complicated shape of $N(M)$ is 
by assuming the existence of two separate galaxy populations 
within halos, central and satellite galaxies. 
%This choice is motivated by reasons based on hydrodynamic simulations 
%\citep{Ber03} and on studies of observed galaxy clusters and groups, 
%which take the brightest cluster galaxies (BCGs) as a different population 
%from the rest of the cluster galaxies. 
This method has been used extensively to interpret galaxy CFs 
\citep{Ham04,Tin05, Phl06, Zhe07, Zeh10, Zhe09}
to constrain how various galaxy samples are distributed among DMH 
as well as whether these galaxies occupy the centers of the DMHs 
or are satellite galaxies \citep{Kra04, Zhe05, Zeh05, Rich12}.
These two populations can be modeled with an HOD described
by a step function above a halo mass limit for central galaxies,
and a power-law for satellite galaxies.
%These two populations can be modeled 
%with a central galaxy HOD described by a 
%step function above a halo mass limit and a power-law for 
%satellite galaxies. 
%In this way the HOD becomes a measure of the combined 
%probability that a halo of mass $M$ hosts a central galaxy and a given 
%number $N_s$ of satellite galaxies.

Similarly, the problem of discussing the abundance and spatial 
distribution of AGN can be reduced to studying how they populate 
their host halos. In fact the observed departure of the 
AGN CF from a power law on small scales (1-2 \mpch) can be physically 
interpreted in the language of the halo model, as the transition between two scales - 
from small scales lying within the DMHs (1-halo term) to those larger than the halo 
(2-halo term). The 1-halo term constrains the  
HOD of satellite AGN and gives us the average profile of pairs
of AGN in groups and clusters of galaxies. The 2-halo term reflects 
the large-scale AGN bias driven by the typical mass of the hosting halos.

Due to the low number density of AGN, there have been 
few results in the literature studying the AGN correlation function
using HOD modeling. Previous works of \citet{Pad09} at $z<0.6$ and 
\citet{Shen10} at $z=3-4$ on QSO using the HOD modeling 
found that $>$25\% and $\geqslant$10\% of their QSOs, 
respectively, are satellites.
\citet{Miy11} described for the first time the shape of the HOD 
of X-ray selected AGN. By using the cross correlation function of 
ROSAT-RASS AGN with SDSS galaxies, they modelled the mean 
AGN occupation of DMHs suggesting that the satellite AGN fraction increases 
slow (or may even decrease) with $M_h$, in contrast with the satellite 
HOD of luminosity-limited samples of galaxies.
%, which, in contrast, 
%grow approximately as $N_s= M_h^{\alpha}$ with $\alpha \sim 1$.\\
Cosmological hydrodynamic simulations have been performed in \citet{Cha12}
to study the mean occupation function of low-luminosity AGN as function
of redshift and luminosity. They used a softened step function 
for the central component plus a rolling-off
power-law for the satellite component with $\alpha=0.3-1.4$ depending
on the redshift and AGN luminosity. Their results suggest
a strong evolution of the AGN occupancy in the redshift
range $z=1-3$ estimated at  three different luminosities
$L_{BOL} \geq 10^{38},10^{40},10^{42}$ s$^{-1}$erg.
\citet{Rich12} modelled the HOD of SDSS quasars at $z \sim 1.4$ 
following this parametrization. They found that the 
satellite occupation becomes 
significant at mass $\sim 10^{14}$\hmsun, i.e. only
the most massive halos host multiple quasars at this redshift 
and only a small satellite fraction ($f_{sat}=7.4 \pm 1.3 \times 10^{-4}$)
of SDSS quasars is required to fit the clustering signal at small scales.
Moreover, they measured that the quasars HOD steepens considerably 
going from z=1.4 to 3.2 over halo mass scales $10^{13-14}$ \hmsun\
and that the characteristic halo mass increases with z for central quasars.
%Recently \citet{Cha12} used cosmological hydrodynamic simulation to
%constrain the AGN HOD of low luminosity AGN as a function of luminosity
%and redshift, showing that the mean occupation function
%can be modelled as a softened step function and an increasing power-law for the 
%central and satellite population, respectively. \citet{Rich12} used these
%results to model the HOD of SDSS quasars at z=1.4, showing that
%at this redshift typically only the most massive halos ($> 10^{14}h^{-1}M_{\odot}$)
%host satellite quasars. 

Despite the diverse methods for studying the HOD, 
counting the number of AGN within 
galaxy groups can constrain quite directly the
average AGN number within a 
halo as a function of halo mass.
The total mass of galaxy groups can be estimated 
via gravitational lensing and the distribution of
AGN within halos can be investigated 
in groups by means of the distribution of the AGN host galaxies. 
Separating the contribution to the occupation of halos
from AGN in satellite or central galaxies can advance
our understanding of the co-evolution
AGN/galaxy and is related to the mechanism of AGN activation.
%In fact, the fact that AGN preferentially reside in central galaxies
%or avoid satellite ones \citep{Pad09, Sta10} provides
%The fact that AGN might preferentially reside in central galaxies
%rather than satellites can advance our understanding 
%of the triggering processes that drive AGN accretion. 

On the other hand, the cross-correlation (CCF) of AGN with galaxy groups
provides additional information about how galaxies and 
BH co-evolve in dense environments. 
In fact the physical processes that drive galaxy
evolution, such as the available cold gas to fuel star formation and the BH
growth, are substantially different in groups and clusters compared to the field.
Many studies over the past decade have presented an evidence that AGN at
$z\sim1$ are more frequently found in groups compared to galaxies
(Georgakakis et al. 2008, Arnold et al. 2009). X-ray observations reveal
that a significant fraction of high-z clusters show overdensities of AGNs in
their outskirts (Henry et al. 1991, Cappi et al.  2001, Ruderman et al.
2005, Cappelluti et al. 2005). 
\begin{figure*}
\plottwo{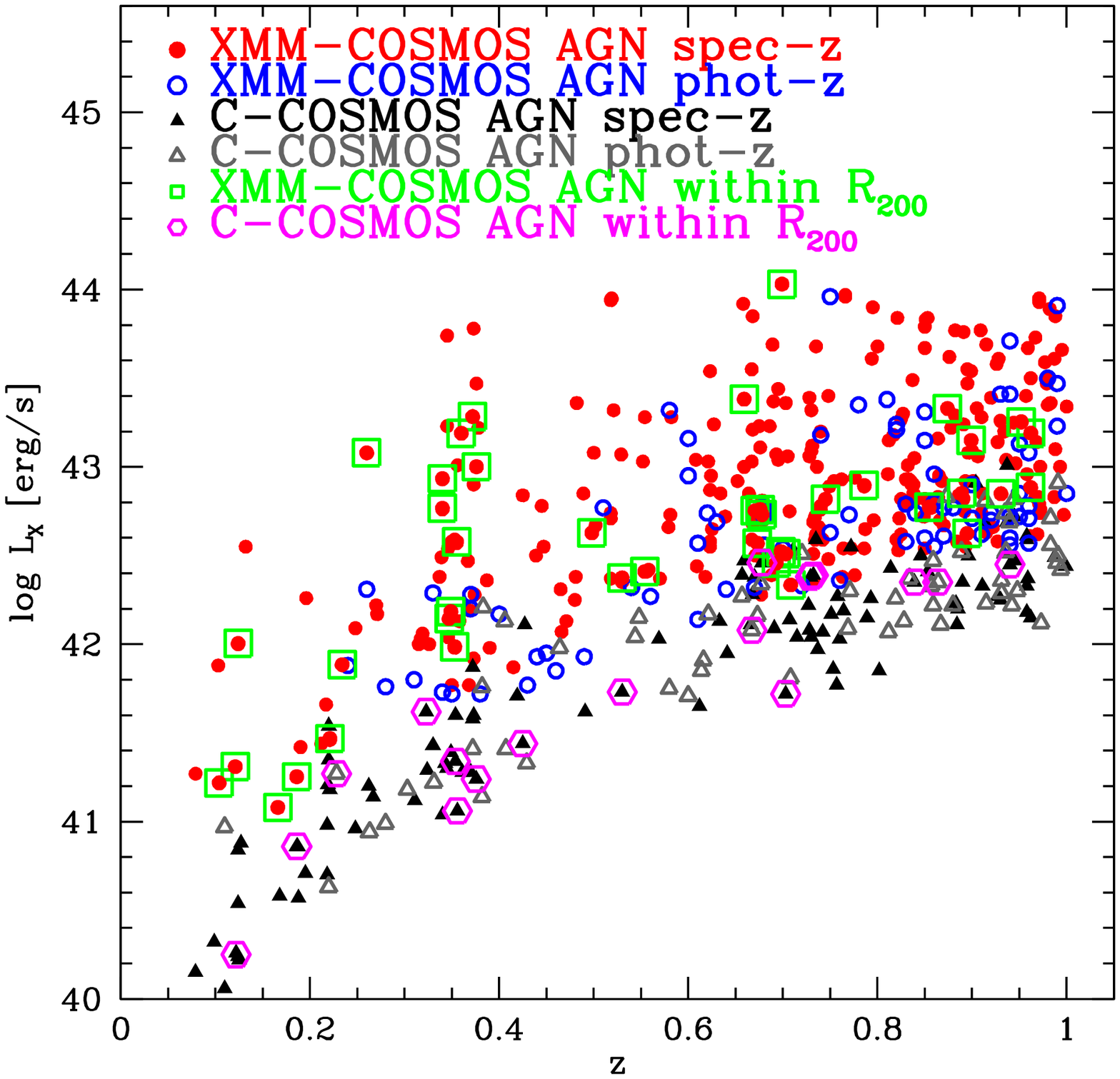}{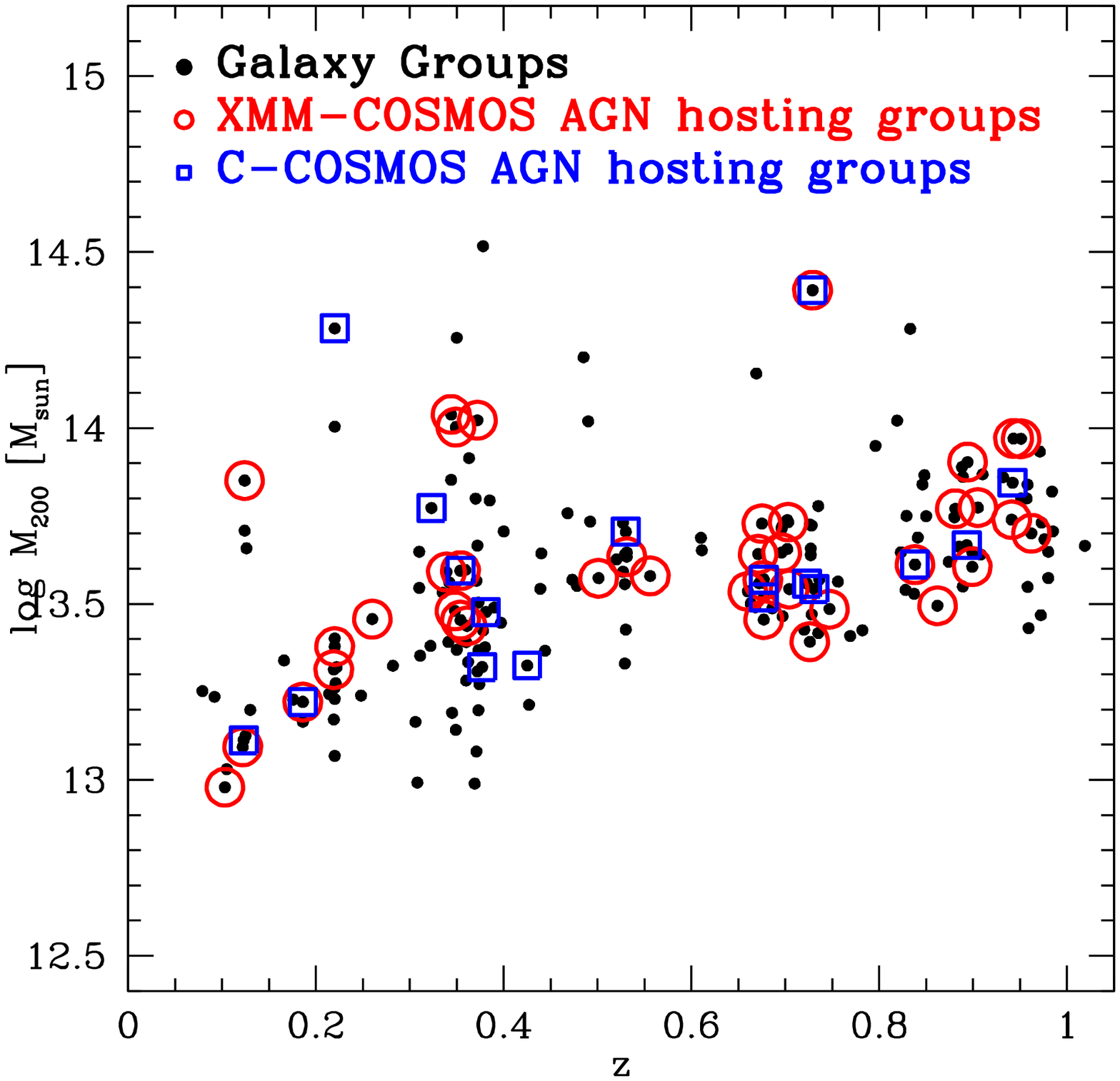}
\caption{\footnotesize \emph{Left panel}: X-ray rest-frame soft luminosity $L_X$ as a function of redshift for XMM-AGN (circles) and C-COSMOS selected AGN (triangle) with know spec-z (filled) and or phot-z (open). The open green squares indicate XMM-AGN within galaxy groups, while the open magenta hexagons represent 17 additional C-COSMOS slected AGN in galaxy groups. \emph{Right panel}: Galaxy group masses $M_{200}$ as a function of redshift for the whole galaxy group sample (black circles) while  open circles (squares) indicate XMM-AGN (C-COSMOS AGN) hosting groups. The mass estimates are defined with respect to 200 times the mean density, in units of $M_{\odot}$.}
\label{fig:histoz}
\end{figure*} 

In this paper, we perform the first direct measurement of 
the mean halo occupation of X-ray AGN and 
of the projected cross-correlation function 
of AGN with galaxy groups,
using a sample of X-ray selected AGN and galaxy groups in the COSMOS field at $z \leq 1$,
from XMM and \textit{Chandra} data.
 %based on the total mass 
%function of X-ray galaxy groups hosting AGN. \\
We use a $\Lambda$CDM cosmology with
$\Omega_M=0.28$, $\Omega_\Lambda=0.72$, $\Omega_b=0.045$, $\sigma_8=0.8$.
For comparison with previous measurements we refer 
to correlation lengths and distances in units of $h^{-1}$ Mpc comoving, 
where $H_0 = 100$ h km s$^{-1}$ Mpc$^{-1}$. AGN luminosities 
and galaxy groups masses are calculated using $h$ = 0.72.
%\begin{figure*}
%\plottwo{fig2a.eps} {fig2b.eps}
%\caption{\footnotesize}
%\label{fig:}
%\end{figure*} 

\section{The Catalogs}\label{sec:cat}

\subsection{AGN Catalog}\label{subsec:AGN}

The \emph{Cosmic Evolution Survey} (COSMOS) is a multiwavelength
survey over $1.4 \times 1.4$ deg$^2$ of equatorial sky
observed by the most advanced astronomical facilities 
HST \citep{Sco07}, SUBARU \citep{Tan07}, Spitzer \citep{San07},
GALEX \citep{Zam07}, \textit{XMM-Newton} \citep{Has07,Cap07, Cap09},
\textit{Chandra} \citep{Elv09, Puc09},
%aimed to study galaxies, large scale structure of the Universe and their co-evolution. \\
designed to thoroughly probe the evolution of galaxies, AGN, 
and dark matter in the context of their cosmic environment (LSS).
%In addiction spectroscopic campaigns have been carried out
%with VIMOS/VLT and extensive spectroscopic follow-up have been granted
%with the IMACS/Magellan, MMT and DEIMOS/KeckII projects.\\
XMM-\emph{Newton} surveyed $2.13$ deg$^2$ of
this field in the 0.5-10 keV energy band for a total of
$\sim$ 1.55 Ms providing a large sample of 1822 point-like
X-ray sources \citep{Cap09}, down to limiting fluxes of $\sim 5 \times 10^{-16}$,
$\sim 3 \times 10^{-15}$ and $\sim 7 \times 10^{-15}$ erg cm$^{-2}$s$^{-1}$
in the soft, hard and ultra hard band, respectively.

The inner part of the COSMOS field ($\sim 0.92$ deg$^2$) has been
imaged for a total of 1.8 Ms by \textit{Chandra} down to limiting fluxes
3-4 times deeper than XMM-COSMOS. Of the 1822 XMM sources, 945
have been observed by \textit{Chandra} and 875 of them are
presented in the C-COSMOS point-like source catalog \citep{Elv09, Puc09, Civ12}.
\citet{Bru10} presented the XMM-COSMOS multiwavelength catalog
of X-ray sources with optical/near-infrared identification, 
multiwavelength properties and redshift information.
Starting from this catalog, we restricted the analysis to a sample
of X-ray AGN (we removed normal galaxies and ambiguous sources)
detected in the soft band which guarantees
the largest sample of X-ray AGN in the COSMOS field, compared to
AGN observed in the hard or ultra-hard band. Moreover,
the selection in only one band allows a more simple
treatment of the AGN X-ray luminosity function used to correct the AGN HOD (\S\ref{sec:HOD})
and of the XMM-and C-COSMOS sensitivity maps 
used to generate the AGN random catalog (\S\ref{sec:CCF}).

Specifically, we selected a sample of 280 and 83 soft XMM-COSMOS AGN with 
spectroscopic and
photometric redshift \citep{Sal11} $z \leq 1$, respectively. 
The use of photometric redshifts for group membership assignment 
has been successfully demonstrated in \citet{Geo11}. 
Note that 184/363 sources are also \textit{Chandra} detected AGN.
%We used the \textit{XMM-Newton} COSMOS survey catalog	\citep{Has07,Cap07, Cap09}, 
%which consists of 1822 bright X-ray sources with accurate identification of their 
%optical counterparts, multiwavelength properties and spectroscopic information as described
%in \citet{Bru10}.
%Specifically, we selected a sample of 280 soft X-ray selected AGN
%with spectroscopic redshift $z\leq1$, which corresponds to the redshift
%range for galaxy groups with highest spectroscopic completeness. 
%We restricted the analysis to AGN detected in the 0.5-2 keV band
%for reasons related to the generation of the random catalog in 
%clustering analysis and to the correction factors applied to the AGN HOD.
%In order to increase the statistics, we included in the direct AGN HOD analysis 
%83 XMM-COSMOS AGN with known photometric redshifts \citep{Sal11}. 
%This has been justified by the photo-z membership assignment shown
%in \citet{Geo11}.
%Note that we made use of this additional AGN sample only in modelling the AGN HOD.
%In fact, when we include AGN with photometric redshift in the clustering 
%analysis, the integral describing the 
%projected correlation function (eq. \ref{eq:wrp})
%never converges.
%We used this AGN catalog with know speczs to estimate the CCF between galaxy groups and COSMOS-AGN
%as described in \s\ref{sec:CCF}. 
In order to test if the AGN halo occupation significantly changes
including sources from the C-COSMOS catalog that are only Chandra detected (hereafter C-COSMOS AGN)
(see \S \ref{sec:HOD}), we included in the 
analysis a sample of 107 and 61 AGN,
detected in the soft band, with known spectroscopic or photometric redshifts
$z \leq 1$, respectively. 
The rest-frame soft X-ray luminosity 
as a function of redshift is shown in Fig. \ref{fig:histoz} (\textit{Left Panel}) for 
XMM-COSMOS AGN (circles) and C-COSMOS AGN (triangles).
%with know spec-z (red circles) or phot-z (blue circles) and for 

%\begin{figure*}
%\plotone{fig5.eps}
%\caption{\footnotesize \emph{Left panel}:}
%\label{fig:HOD}
%\end{figure*} 

\subsection{Galaxy Group Catalog}\label{subsec:groups}

We used a catalog with 270 galaxy groups, detected in the co-added \textit{XMM-Newton}
and \textit{Chandra} data.
The general data reduction process is described in \citet{Fin07} and 
details regarding improvements and modification to 
the initial catalog are given in \citet{Lea10} and \citet{Geo11}.
The identification of the groups has been done using the red sequence technique,
and spectroscopic identification of
groups has been achieved through zCOSMOS-BRIGHT program \citep{Lil09},
targeted follow-up using IMACS/Magellan and
FORS2/VLT \citep{Geo11}, Gemini GMOS-S \citep{Bal11} as well as through secondary targets
on Keck runs by COSMOS collaboration. At $z \leq1$ the spectroscopic
completeness has achieved 90\% (A. Finoguenov 2012, in preparation).

As described in \citet{Lea10}, the total X-ray fluxes have been obtained 
from the measured fluxes by assuming a beta profile and by 
removing the flux that is due to embedded AGN point sources.
Some of the faint Chandra AGN could not be removed
from the group fluxes, with their contribution to the total flux being $<10\%$.
%the total X-ray fluxes are obtained
%from the residual images after the point-source subtraction. The aperture of
%the flux extraction has been determined by the statistical significance of
%X-ray flux and a correction for the missing flux has been done by assuming a
%beta profile, as in \citet{Fin07}. 
The rest-frame luminosities have been
computed in \citet{Fin07} and \citet{Lea10} from the total flux following $L_{0.1-2.4 keV}=4 \pi
d_{L}^2K(z,T)C_{\beta}(z,T)F_d$, where $K(z,T)$ is the K-correction and
$C_{\beta}(z,T)$ is an iterative correction factor, while to estimate
the temperature of each group they used the $L_X-T$
relation of \citet{Mar98}.

A quality flag (hereafter 'XFLAG') is assigned 
to the reliability of the optical counterpart, with flags 1 and 2 indicating 
a secure association, and higher flags indicating potential problems 
due to projections with other sources or bad photometry due to bright stars 
in the foreground. In detail, XFLAG=1,2 are assigned to groups with a 
confident spectroscopic association, while systems with only the 
red sequence identification have \emph{XFLAG=3} \citep[see][for more details]{Lea10}.
%\emph{XFLAG=1} is assigned to groups with a single
%optical counterpart and with a clear X-ray peak. The flag \emph{XFLAG=2} is
%assigned to systems for which the extended X-ray emission is subject to
%projection effects, but for which the various projections can be
%disentangled. Systems with only the red sequence identification are assigned
%\emph{XFLAG=3}. \emph{XFLAG=4} indicates that there are several equally
%possible optical counterparts and \emph{XFLAG=5} is assigned to systems for
%which the optical counterpart is uncertain  

The line-of-sight position of the group is assigned to be
the centroid of the X-ray emission (the accuracy of the determination of the
X-ray center is higher for \emph{XFLAG=1}). 
If the X-ray centroid is not
precise enough to be used directly,the Most Massive Central
Galaxy (MMCG) located near the peak of the X-ray emission has been used to
trace the center of the DM halos of groups \citep[see][for more details]{Lea10, Geo11}. 
Group masses $M_{200}$ are assigned from an empirical mass-luminosity
relation, described in \citet{Lea10},
\begin{eqnarray}
log_{10}(M_{200,c}) = p_0 - log_{10}E(z) + log_{10}(M_{0})\\
+ p_1[log_{10}(L_{x}/E(z)) - log_{10}(L_{0})] \nonumber
\end{eqnarray}
where $M_{200}$ is the mass within the radius containing the
density of matter 200 times the critical density, in units of $M_{\odot}$.
$\left\lbrace p_0,p_1 \right\rbrace =\left\lbrace 0.729538, 0.561657 \right\rbrace $
are the fitting parameters, $\left\lbrace
logM_0,logL_0 \right\rbrace =\left\lbrace 13,42.5\right\rbrace $ are the
calibration parameters and $E(z)$ is the correction for redshift evolution
of scaling relations, which has been shown in \citet{Lea10} to
reproduce well the $L_{X}-M$ relation of COSMOS groups.

In order to be consistent in comparing these mass values with the ones
obtained studying the clustering properties of groups, we accounted for the difference between the
mass defined with respect to 200 times the critical density and with respect to 200 times the mean density
(hereafter $M_{200}$ refers to masses obtained using the definition with
respect to the mean density). In fact, the absolute bias which we are going
to derive from the DM correlation function is based on the shape of the DM
mass function defined with respect to the mean density. 
%and is derived for the mass determined with respect to the
%mean density.  
Starting from the Navarro-Frenk-White (NFW) profile with a concentration parameter
$c=5$, we derived the relation between the two mass definitions,
$M_{200,m}=M_{200,c} \times \Omega(z)^{-0.134}$.  

In this work, we make use of galaxy groups with $z \leq 1$ and identification flag \emph{$XFLAG\leq 3$}
which removes problematic identification cases,
obtaining a catalog of 189 X-ray galaxy groups 
over 1.64 deg$^2$ with a 
rest-frame 0.1-2.4 keV luminosity	 range of $41.3<log(L_X [s^{-1}$erg]$ )<	44.1$,
and mass range of $13 < logM_{200}[M_{\odot}] < 14.5$
(see Fig. \ref{fig:histoz}, \textit{Right Panel}).

\begin{figure*}
\plottwo{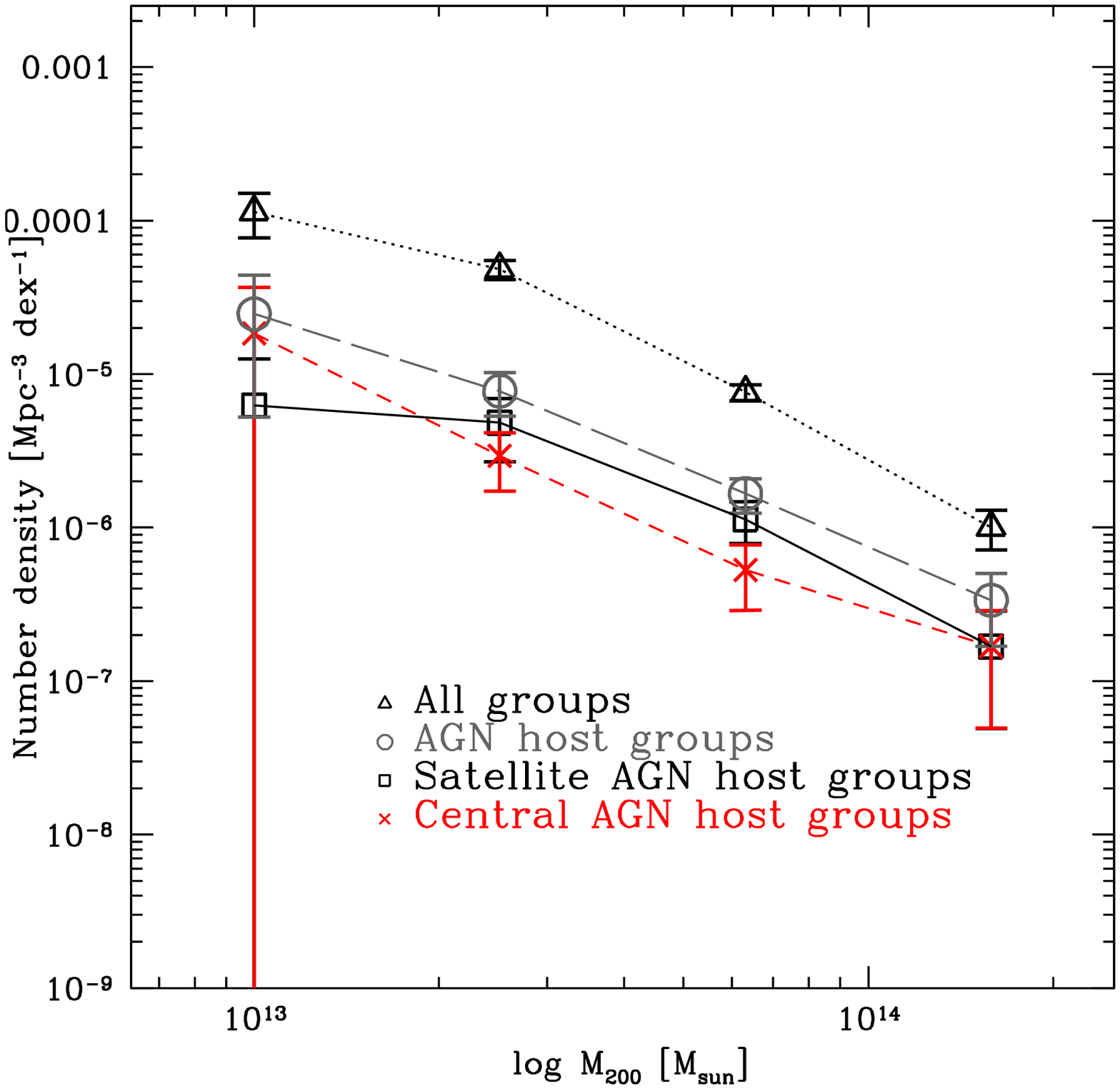}{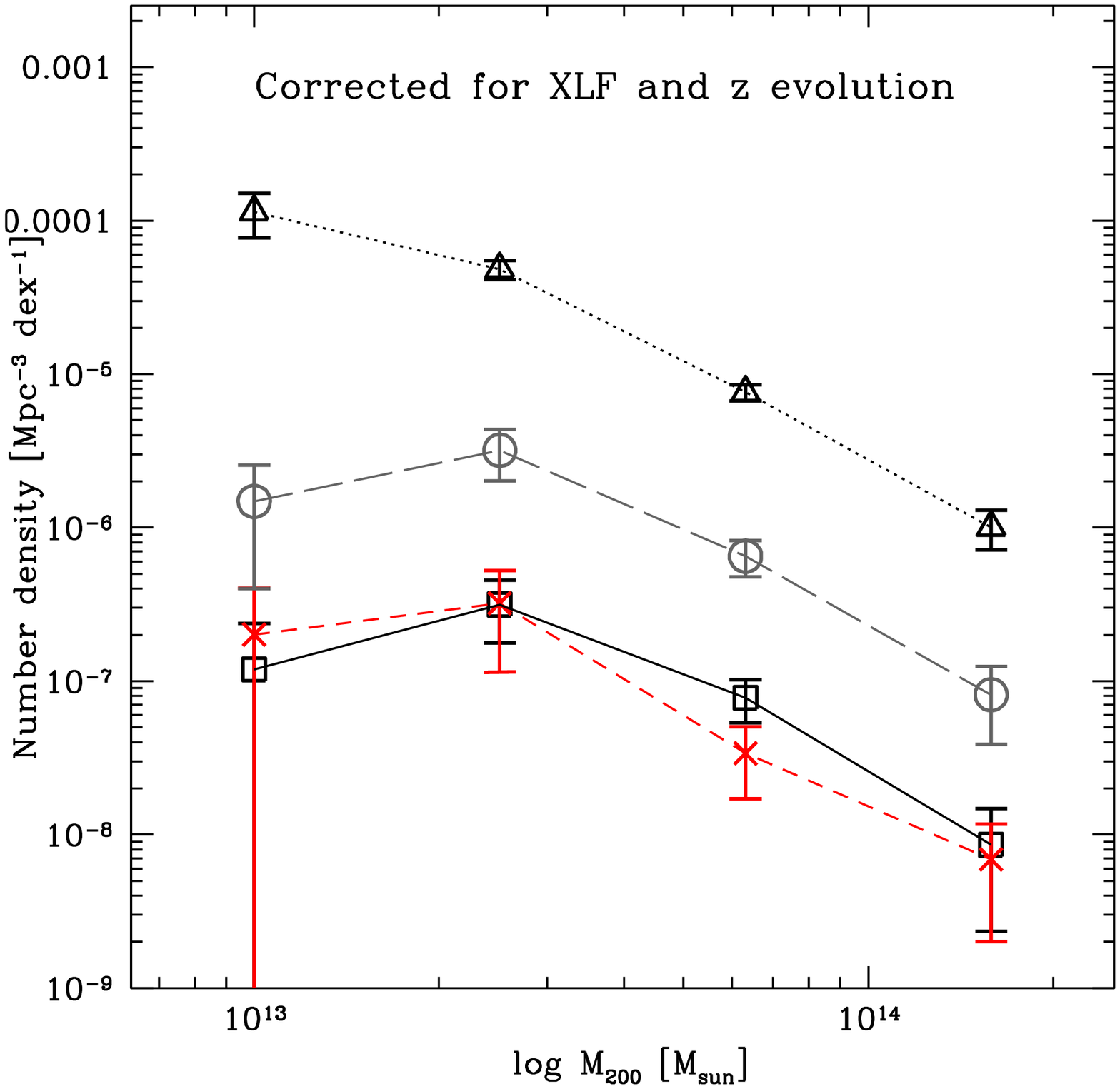}
\caption{\footnotesize \textit{Left Panel}: Mass function of X-ray galaxy groups (triangles) and 
AGN host groups (circles). The black squares (red crosses) show the mass function of groups hosting an AGN
in satellite (central) galaxies. \textit{Right Panel}: Same as left panel when correcting for the AGN soft XLF and the redshift evolution.}
\label{fig:massfunc}
\end{figure*} 

\subsection{AGN in galaxy groups}\label{subsec:AGNingroups}

We define group members as AGN located within $< 3\sigma$ 
and $< R_{200}$ from the group centers, where $\sigma$ is the
group line-of-sight velocity dispersion and $R_{200}$ is the virial radius 
of a group within which the mean density is 200 times 
the mean density of the Universe at the group redshift.
%Using this method starting from the sample of XMM-AGN with know spec-z, 
%we found 27 sources within $R_{200}$. 

In this analysis we used the sample of 363 XMM-COSMOS AGN with $z \leq 1$ described in \S\ref{subsec:AGN}
and we found 41 sources (35/41 are also \textit{Chandra} detected) in 
galaxy groups with median $\langle z \rangle = 0.55$ and median $\langle L_{X} \rangle =10^{42.6}$ $s^{-1}$erg.
%In order to have a complete AGN sample, we selected a subset of XMM-AGN with spectroscopic or photometric 
%$z \leq 1$ and X-ray rest-frame soft luminosity $logL_X>42.5$ erg/s. We set the luminosity cut such that we ensure detection
%of AGN to any redshift and we found 27 XMM-AGN within $R_{200}$ (gray open circles).
%In order to increase the statistics, 
%we estimated the number of AGN in galaxy groups, by making use of 170 Chandra-COSMOS AGN,
%with know spectroscopic or photometric redshifts $z\leq 1$ \citep{Sal11, Puc09} and
%by including all the 300 XMM-AGN without any luminosity cut. 
%This allowed us to collect a sample of 58 COSMOS-AGN in 
%galaxy groups with $z\leq1$ (green open squares): 17 from the Chandra catalog and 41 from the XMM sample.
%Fig. \ref{fig:histoz}, right panel, shows the redshift distribution of
%XMM (black) and Chandra (grey) AGN within $R_{200}$.
%In \S\ref{sec:HOD} we measure the difference in the AGN HOD 
%we compare the AGN HOD estimated by 
%using 41 XMM-AGN in galaxy groups with the HOD 
When we include in the analysis the sample of 168 C-COSMOS AGN (\S\ref{subsec:AGN}),
we found 17 additional
AGN in galaxy groups, with known spectroscopic or
photometric $z \leq 1$.
As expected C-COSMOS AGN have lower soft fluxes respect to XMM-COSMOS AGN
at any redshift, with a median $L_{X}=10^{41.7}$ $s^{-1}$erg.
In particular we found 2 galaxy groups at $z \sim 0.1$ and 2 groups at
$z \sim 0.7-0.8$ with 2 AGN per halo and 1 group with 3 AGN at $z \sim 0.35$.
All the properties of AGN and galaxy group samples are
summarized in Table \ref{tbl-1}, while 
Table \ref{tbl-2} shows the catalog of 58 XMM + C-COSMOS AGN in galaxy groups.
%structured as col (1): galaxy groups ID, col (2,3): galaxy groups $\alpha$ and $\delta$, col (4):
%photometric or spectroscopic redshift, col (5): galaxy group mass estimates, col (6):
%AGN ID in XMM-COSMOS catalog, col (7,8): AGN $\alpha$ and $\delta$, col (9): AGN ID in
%C-COSMOS catalog, col (10): spectroscopic or photometric redshift, col (11):
%rest-frame soft X-ray AGN luminosity, col (12): =0 or =1 for AGN among
%satellite or central galaxies, respectively.
%Note that we made use of XMM and Chandra AGN with phot-zs only in modelling the AGN HOD.
%In fact, we estimated the cross-correlation function AGN-galaxy groups by using XMM-AGN with
%spec-z. When we include phot-zs in clustering analysis, the integral describing the 
%projected correlation function (eq. \ref{eq:wrp})
%never converges.
For the sources with known spectroscopic redshifts (47/58)
we know the classification in BL or non-BL AGN as described
in \citet{Bru10} and \citet{Civ12} for XMM and C-COSMOS AGN,
respectively. In detail, we found 43 non-BL and 4 BL 
in the XMM + C-COSMOS AGN sample.
Fig. \ref{fig:histoz} (\textit{Left Panel}) shows the rest-frame soft 
X-ray luminosity as a function of redshift 
%XMM-COSMOS AGN with known spec-z (red circles), phot-z (blue circles)
for the subsamples of XMM (green squares) and 
C-COSMOS selected AGN (magenta hexagon) within $R_{200}$.

We cross-matched the sample of AGN in groups
with a galaxy membership catalog \citep{Lea07,Geo11}
and we verified that AGN classified as group members based on our method,
have host galaxies associated with the same galaxy groups. 
%(most of them with a probability $>0.5$).\\
We divided the sample of 58 AGN in groups in two subsets, 
according to their association with BCGs.
In detail we found that 22/58 (16/41) AGN are in central galaxies, 
while 36/58 (25/41) are in satellites.

\begin{deluxetable}{llll}
\tabletypesize{\scriptsize}
\tablewidth{0pt}
\tablecaption{Properties of the Group and AGN Samples \label{tbl-1}}
\tablehead{
\colhead{(1)} &
\colhead{(2)} &
\colhead{(3)} &
\colhead{(4)} \\
\colhead{Sample} &
\colhead{N} & 
\colhead{$\langle z \rangle$\tablenotemark{a}} &
\colhead{$\langle L_{X} \rangle$\tablenotemark{b}} \\
\colhead{} &
\colhead{} &
\colhead{} &
\colhead{} }
\startdata
\multicolumn{4}{l}{}\\
XMM-AGN  & 363  & 0.66  & $10^{42.8}$ \\
C-COSMOS AGN & 168 & 0.56 & $10^{41.8}$  \\
XMM-AGN in $R_{200}$ & 41 & 0.55 & $10^{42.6}$ \\
C-COSMOS AGN in $R_{200}$ & 17 & 0.53 & $10^{41.7}$ \\
XMM-AGN in the field\tablenotemark{d} & 253 & 0.67 \tablenotemark{e} & $10^{42.8}$ \\  
XMM+C-COSMOS AGN in $R_{200}$ & 58 & 0.55 & $10^{42.3}$ \\
- Satellites & 36 & 0.56 & $10^{42.2}$ \\
- Centrals & 22 &  0.50 & $10^{42.4}$ \\
\multicolumn{4}{l}{}\\
& & & $\langle M_{200} \rangle$\tablenotemark{c} \\
& & &  \\
Galaxy Groups & 189 & 0.56 & $10^{13.60}$  \\
XMM+C-COSMOS AGN host groups & 52 & 0.55 & $10^{13.62}$ \\
\enddata
\tablenotetext{a}{Median redshift of the sample.}
\tablenotetext{b}{In units of $h_{70}^{2}$ erg $s^{-1}$}
\tablenotetext{c}{Mass defined respect to 200 times the mean density known with a 20\% error, in units of $M_{\odot}$.}
\tablenotetext{d}{AGN sample used in estimating the CCF.}
\tablenotetext{e}{Only spectroscopic redshifts.}
\end{deluxetable}

\section{Halo Mass Function and AGN HOD}\label{sec:HOD}

Fig. \ref{fig:massfunc} (\textit{Left Panel}) shows the mass function of all 
X-ray galaxy groups and those marked by AGN presence,
showing separately the contributions of groups hosting an AGN in central or
satellite galaxies. 
We calculated the mass function by using the standard $1/V_{max}$
method (Schmidt 1968) and we counted twice galaxy
groups with 2 AGN.
Hence, in the $i^{th}$ mass bin, the comoving space density
($n_i$) and its corresponding error ($\sigma_i$) are computed by (see Bondi et al. 2008):
\begin{align}
n_i  = \sum_{j} \frac{1}{V_{max}^j}  && 
\sigma_i  = \sum_{j} \sqrt{\left( \frac{1}{V_{max}^j} \right)^2}
\end{align}
In estimating the average number of AGN occupying a halo
of mass $M_{200}$, some major effects need to be taken into 
consideration. The sample of AGN in $R_{200}$
is a flux-limited sample and brighter AGN are detected at higher redshift.
Similarly, we observe galaxy groups with small halo mass only at low redshift.
However the relatively small number of COSMOS AGN in 
galaxy groups does not allow us to select a volume 
complete subsample by using a cut in luminosity at log$L_{X}$[\ergs]=42.4.
%or AGN subsamples with different luminosity thresholds as a
%function of redshift. 
We have therefore made a correction for the effect of changes
in the AGN density as a function of redshift and limiting luminosity,
by using the AGN X-ray luminosity function (XLF).
%We assumed that the shape of the XLF of AGN in galaxy groups 
%is the same as that of the total AGN population,
%apart from the different normalization.
It has been shown in different works that
the Luminosity Dependent Density Evolution (LDDE) model for the XLF provides the best framework that describes
the evolutionary properties of AGN, both in the soft \citep{Miy00, Has05}
and hard X-rays \citep{Ued03, LaF05}. 
In this paper we modelled the AGN soft XLF
with the LDDE XLF described in \citet{Ebr09},
because they modelled
the soft XLF including in the analysis both type 1 and type 2 AGN,
but we verified that using different best-fit parameters of the
global XLF expression, the resulting mean AGN occupation
stays within the error bars. 

\begin{figure*}
\plottwo{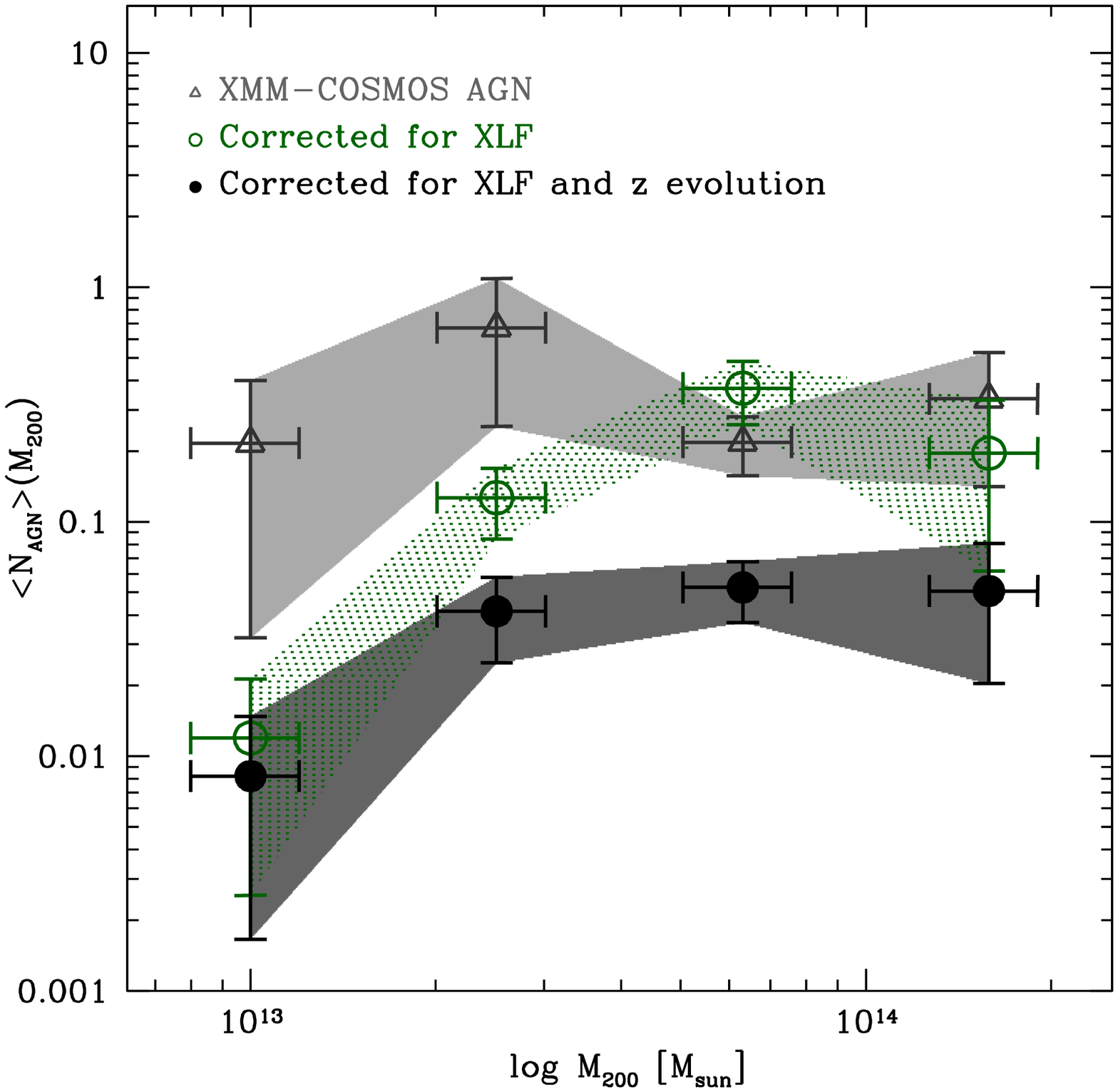}{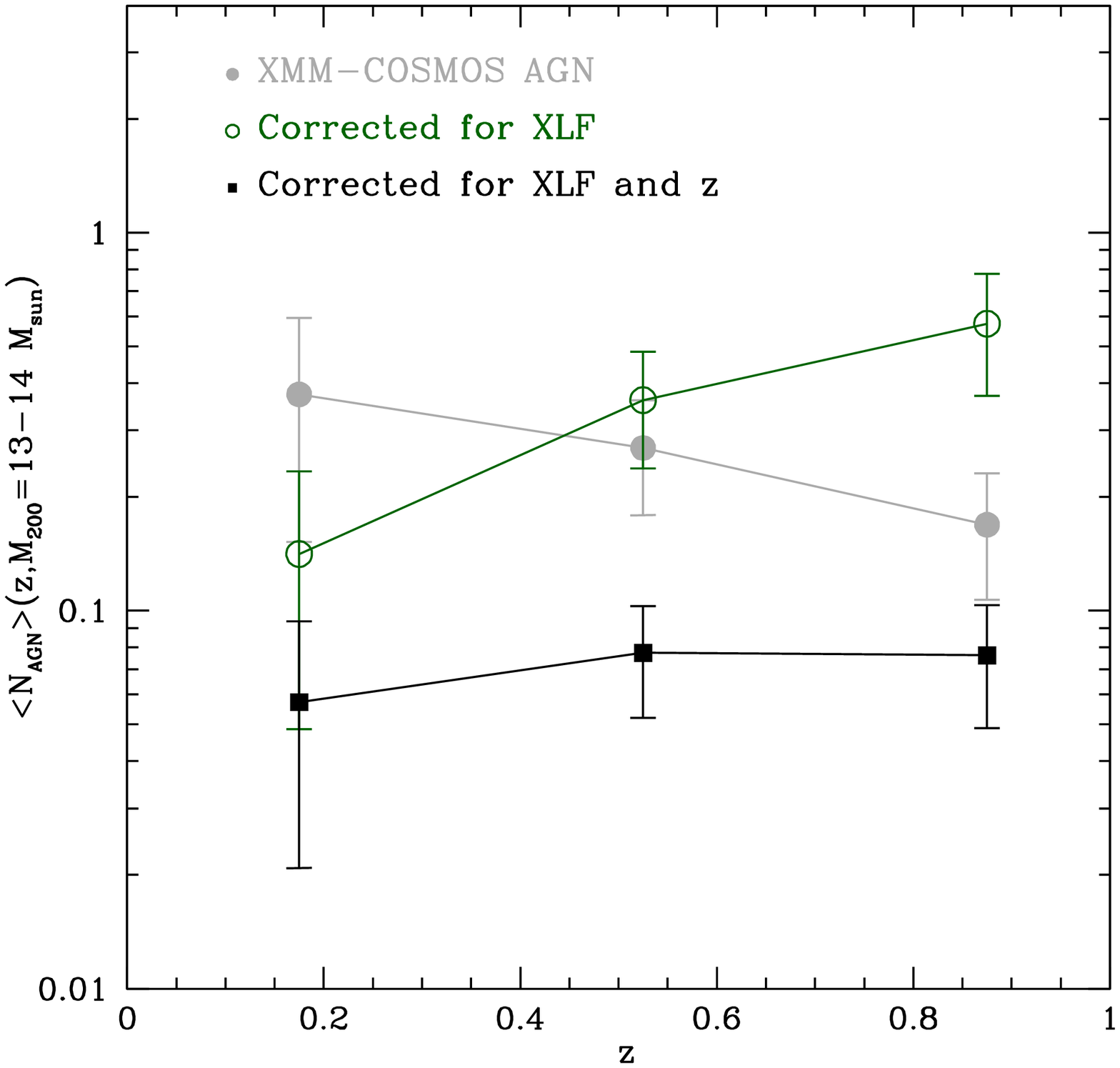}
\caption{\footnotesize \textit{Left Panel}: Observed occupation of galaxy groups by XMM-COSMOS AGN as a function of the halo 
mass (open grey triangles) and after correcting for the soft XLF (green open circles) and for both the XLF
 and the AGN redshift evolution. \textit{Right Panel}: Mean AGN occupation as a function of redshift for AGN hosting groups with $M_{200}=13-14M_{\odot}$ (colors with same meaning as left panel.)}
\label{fig:HODagn}
\end{figure*}

Then for each AGN redshift we defined two weights: $w$ to
correct for the fact that we are including in the analysis AGN with 
log$L_X$[\ergs]$<$42.2 and $w_0$ to correct for this effect plus the redshift evolution of the AGN density:
\begin{equation}\label{eq:w}
w(z) = \frac{\int_{42.4}^{\infty} \phi(z,L_X) dL_X } {\int_{L_{lim}(z)}^{\infty} \phi(z,L_X)  dL_X}
\end{equation}
\begin{equation}\label{eq:w0}
w_0(z) = \frac{\int_{42.4}^{\infty} \phi(z=0,L_X) dL_X } {\int_{L_{lim}(z)}^{\infty} \phi(z,L_X)  dL_X}
\end{equation}
$\phi(z,L_X)$ is the soft XLF proposed in \cite{Ebr09}
and $L_{lim}$ depends on the survey flux limit and is a function of redshift.
In detail, following Equations 9,11 and 12 and Table 2
in \citet{Ebr09}, we described the shape of the present-day luminosity function 
with slopes $\gamma_1=0.72 \pm 0.02$ and $\gamma_2=2.04 \pm 0.04$, 
log$L_0$[h$_{70}$\ergs] = 43.65 $\pm$ 0.05 which is the value of the luminosity where the change of slope occurs 
and normalization $A=3.76 \pm 0.38 \times 10^{-6}$h$_{70}^{3}$Mpc$^{-3}$. We estimated the evolution factor
assuming $p_1=3.38 \pm 0.09$, $p_2=-1.5$, $z_c=1.42$, log$L_a$ = 44.6h$_{70}$\ergs\ and $\alpha=0.100 \pm 0.005$.

Based on Eq. \ref{eq:w} and \ref{eq:w0}, the comoving space
density $n_i$ of AGN hosting groups corrected for the XLF is given by:
\begin{equation}\label{eq:nw}
n_i = \sum_{j} \frac{w^j(z,L_{X}) }{V_{max,j}}
\end{equation}
while the $n_i$ corrected for both the XLF and the z evolution is estimated by using: 
\begin{equation}\label{eq:nw0}
n_i = \sum_{j} \frac{w_0^j(z=0,L_{X}) }{V_{max,j}}
\end{equation}
%Then in each mass bin, 
%In detail, we weighted each observed AGN in each halo mass bin,
%by the factor $N_{obs}/N_{XLF}$, where $N_{obs}$ is the number of
%AGN in galaxy groups found in the luminosity-redshift bin of the particular
%AGN and $N_{XLF}$ is normalized to the observed number of AGN 
%in galaxy groups at $z=0$. 
%\textbf{The observed number of AGN in galaxy
%groups in different luminosity and redshift bins follows the trend
%predicted by the XLF, i.e. at higher luminosity the AGN density
%peaks at higher redshift compared with low luminosity AGN.
%This supports our assumption that the shape of the XLF of AGN in galaxy groups 
%is the same as that of the total AGN population,
%apart from the different normalization.}
%Moreover, we corrected for the intrinsic redshift evolution of the AGN density  
%by using the density evolution factor $e_d(z,L_X)$ \citep{Ebr09}, which describes
%the evolution of the AGN XLF with redshift accordingly with the luminosity.
Fig. \ref{fig:massfunc} (\textit{Right Panel}) shows the mass function
when all these effects are corrected.

The ratio between the mass function of X-ray groups hosting 
AGN within $R_{200}$ by that of all X-ray galaxy groups
generates the mean AGN HOD, which describes 
the occupation of DM halos by AGN. 
Fig. \ref{fig:HODagn} (\textit{Left Panel})
shows the observed average number of XMM-COSMOS AGN in a halo of given mass as a function
of $M_{200}$ (grey triangles) and the average number after correcting for the XLF (green open circles)
and for both the XLF and the z evolution (black filled circles). 
The redshift evolution of the AGN fraction is shown in Fig. \ref{fig:massfunc} (\textit{Right Panel}).
The grey circles represent the mean AGN occupation as a function of redshift for halos with $M_{200}=13-14M_{\odot}$
while the green circles and the black squares show the AGN fraction when we correct for the
XLF only and for both the XLF and the z evolution effects, respectively.  
When we correct for the XLF, the mean occupation increases with z 
since in the AGN luminosity range of our interest, the AGN density
increases with z, while the redshift correction removes this trend
producing a constant AGN fraction.

%Moreover by dividing the the mass
%function of satellite (central) AGN host groups by that of
%all X-ray galaxy groups,
%we provide the fraction of AGN among satellite (central) 
%galaxies as a function of halo mass (see fig. \ref{fig:HODsatcen}).
%Usually we model the mean AGN occupation function in halos
%by decomposing it into the central and satellite contribution
%$ \langle F_{AGN}(M) \rangle  = \langle F_{cen}(M) \rangle  +  \langle F_{sat}(M) \rangle$
%where $F_{sat}$ and $F_{cen}$ are described by a power-law and a
%step function at $M>M_{min}$ \citep{Miy11}. 
%
%As shown in fig. \ref{fig:HODsatcen}, due to the low statistics,
%we can not constrain the HOD of satellites and central AGN 
%separately (the data points are consistent within the error bars).
We fitted the total AGN HOD assuming the model with a rolling-off power-law:
\begin{equation}\label{Fagn}
\langle N_{AGN} \rangle (M_h) = f_{a} \left(  \frac{M_h}{M_1} \right)^{\alpha} exp \left(  \frac{M_{cut}}{M_h} \right) 
\end{equation}
where $f_{a}$ is the normalization, $M_1$ is the halo mass
at which the number of central AGN is equal to that 
of satellite AGN, $M_{cut}$ is a cut-off mass scale.
With our data alone, we cannot make meaningful constraints on $M_1$,
therefore we fixed log$M_1=13.8M_{\odot}$ following the
results of \citet{Miy11}. We verified that
the result does not change for log$M_1=13-14.2 M_{\odot}$,
due to the fact that the fraction of AGN among central and
satellite galaxies are comparable in this mass range.
For the modelling of the rolling-off power-law,
we assumed log$M_{cut}=13.4M_{\odot}$ which corresponds
to the mass below which our data points decay exponentially.

\begin{figure*}
\plottwo{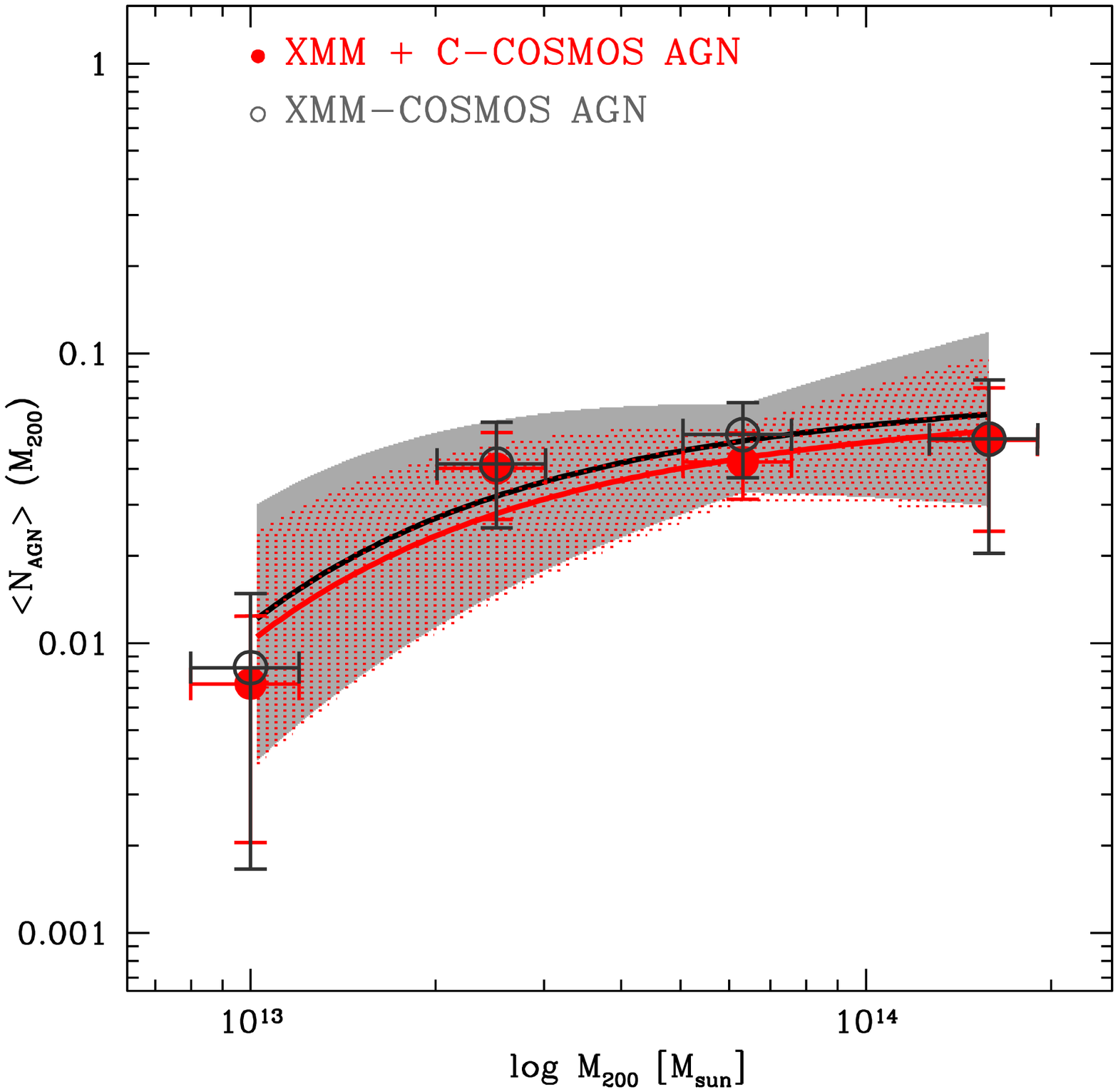}{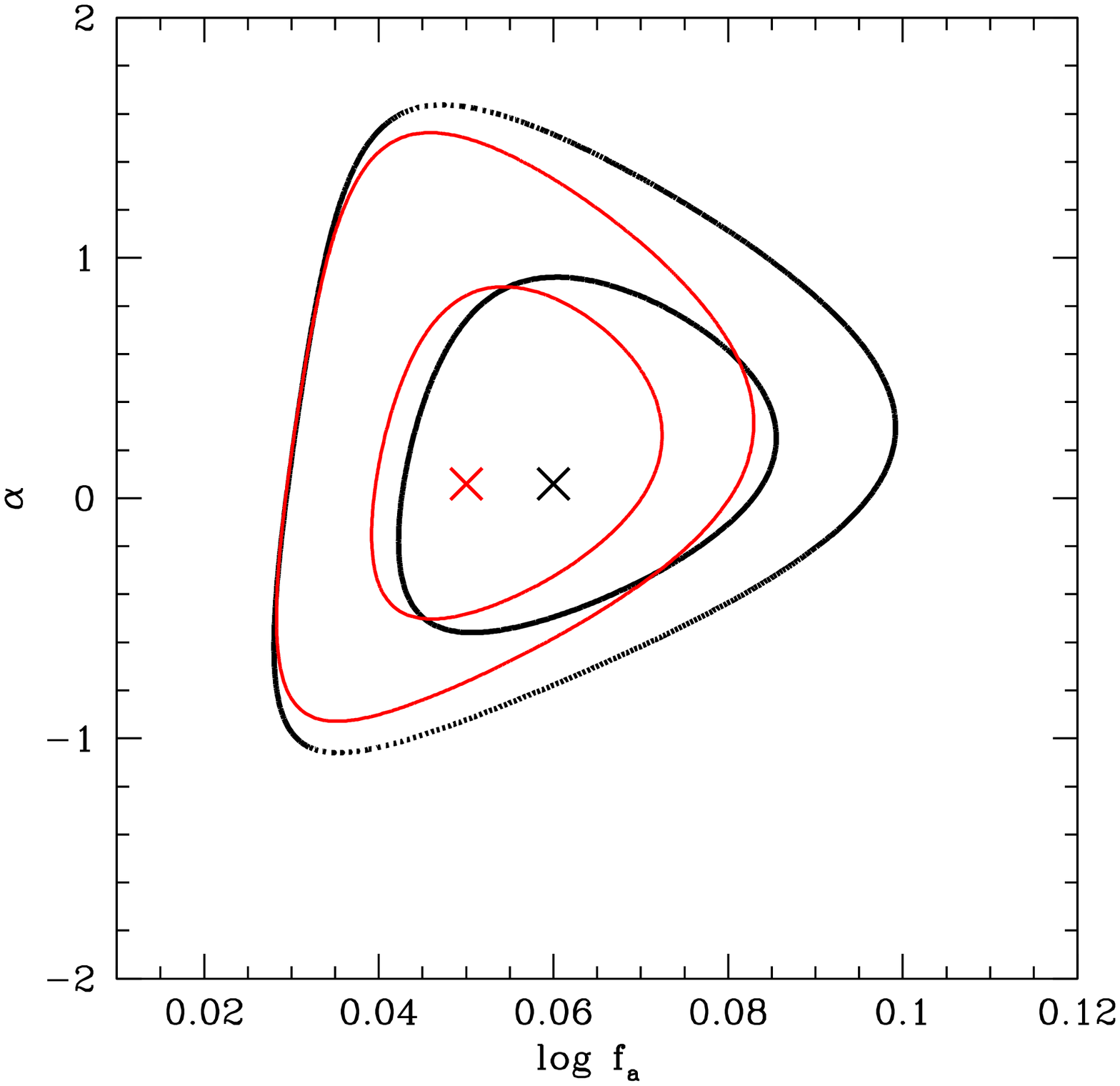}
\caption{\footnotesize \textit{Left Panel}: Occupation of galaxy groups by 41 XMM-COSMOS AGN 
(black open circles) and 58 XMM + C-COSMOS AGN (red filled circles) 
as a function of the halo mass, when correcting for the XLF and for the redshift 
evolution of the AGN density. The fit assuming a rolling-off power-law
dependence of the HOD is shown as the solid black lines (best fit) and shaded regions
($1\sigma$ confidence interval, $\Delta \chi^2=2.3$). \textit{Right Panel}: the confidence contours of the 
power-law best-fit parameters $\alpha$ and $f_a$, for XMM-COSMOS AGN (black) and for 
XMM + C-COSMOS AGN (red) in galaxy groups. The contours mark the 68.3\% and 95.4\% confidence levels 
(respectively corresponding to $\Delta \chi^2$ = 2.3 and 6.17).}
\label{fig:HOD}
\end{figure*}

%In order to better constrain the shape of the AGN halo occupation,
%we included in the analysis 17 additonal Chandra-COSMOS AGN with $z \leq 1$.
%Using the same method described for XMM-AGN,
%we corrected the halo occupation of Chandra/XMM-AGN for the
%redshift evolution and the XLF. 
%We fitted the HOD with Eq. \ref{Fagn} obtaining as best fit parameters
%$\alpha=-0.31(-0.84,0.65)$ and $f_a=0.2(0.12,0.27)$ 
%($1\sigma$ errors) which are consistent within the errors with the ones 
%found for XMM-AGN in galaxy groups. 

We obtained constraints in the ($\alpha$, $f_{a}$) - space and we found 
as best fit parameters of the mean AGN HOD,
$\alpha=0.06(-0.22;+0.45)$ and $f_a=0.06(0.04;0.08)$, 
where the $68 \%$ confidence interval for a combined two parameter
fit ($\Delta \chi^2=2.3$) is given in the brackets.
Fig. \ref{fig:HOD} (\textit{Left Panel}) shows the mean occupation
of XMM-COSMOS AGN with the best fit parameters (solid black line), 
and 1$\sigma$ confidence interval (shaded grey region),
compared to the mean AGN HOD including C-COSMOS AGN in the analysis. 
The solid red line corresponds to the best fit model 
for the mean occupation of XMM+C-COSMOS AGN, with
$\alpha=0.06(-0.22,0.36)$ and $f_a=0.05(0.04,0.06)$,
while the shaded region is the $1\sigma$ confidence interval
(see Fig. \ref{fig:HOD}, \textit{Right Panel}).
%The power-law index $\alpha$ is consistent within the errors
%with the one found for XMM-COSMOS AGN in galaxy groups,
%while the difference in the normalization
%($f_a$) reflects the fact that we are including fainter X-ray AGN
%and then moving towards lower limiting luminosity.}

\begin{figure}
\plotone{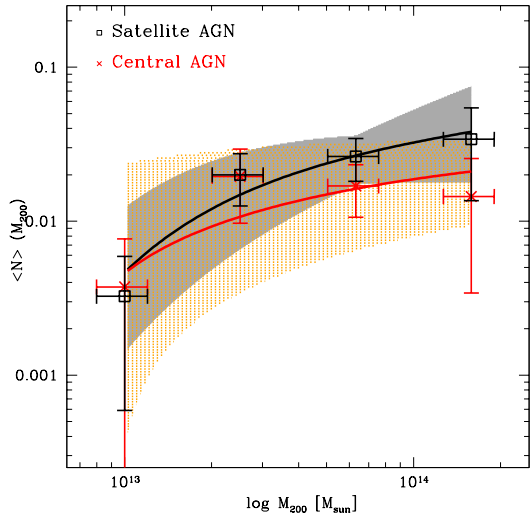}
\caption{\footnotesize Occupation of galaxy groups by satellite AGN (black squares) and
central AGN (red crosses) as a function of the halo mass when correcting for the redshift 
evolution of the AGN density and for the soft XLF. The fit assuming a rolling-off power-law
dependence is shown as solid black line (best fit) and dashed grey region
($1\sigma$ confidence interval) for AGN among satellite galaxies. The HOD for central AGN
has been modelled with a softened step function  (solid red line) where the shaded orange region marks
the 68.3\% confidence level.}
\label{fig:HODsatcen}
\end{figure}

Moreover by dividing the mass
function of satellite (central) AGN host groups by that of
all X-ray galaxy groups,
we provide the fraction of AGN among satellite (central) 
galaxies as a function of halo mass (see Fig. \ref{fig:HODsatcen}).
We model the mean AGN occupation function in halos
by decomposing it into the central and satellite contribution
$ \langle N_{AGN} \rangle  (M_h) = \langle N_{cen} \rangle  (M_h) +  \langle N_{sat} \rangle (M_h) $:
\begin{eqnarray}
\langle N_{cen} \rangle  (M_h) = f_{a}^{\prime}  ~ erf \left(  \frac{logM_h-logM_{min}}{\sigma_{logM}} \right)  \\
\langle N_{sat} \rangle (M_h) = f_{a}^{\prime}  ~ \left(  \frac{M_h}{M_{1}} \right)^{\alpha_s}  exp(-M_{cut}/M_h)
\end{eqnarray}
where the central AGN occupation follows a softened step function 
and the satellite occupation a rolling-off power law (e.g. Kravtsov et a. 2004,
Zheng et al. 2005, Zehavi et al. 2005, Tinker et al. 2005, Conroy et al. 2006, Chatterjee et al. 2011, Richardson et al. 2012).
In this formalism there are four free parameters $f_a^{\prime}$, $M_{min}$, $\sigma_{logM}$ and $\alpha_s$, 
where $M_{min}$ is the minimum mass where the occupation of central AGN is zero.
%where $\langle N_{sat} \rangle$ and $\langle N_{cen} \rangle$ are described by a power-law and a
%step function at $M>M_{min}$ \citep{Miy11} and $f_{a,c}$ is the AGN fraction among central galaxies.

As shown in Fig. \ref{fig:HODsatcen}, 
the HOD of central AGN is described by a softened step function with 
log$M_{min}[M_{\odot}]=12.75(12.10,12.95)$ and $\sigma_{logM}=1.46(0.4,4.0)$
where the errors are the $1\sigma$ confidence intervals estimated
by using the $\chi^2$ minimization technique with 2 free parameters.
On the other hand, the satellite AGN HOD
%decreases with the halo mass
%we can not constrain the HOD of satellites and central AGN 
%separately (the data points are consistent within the error bars).
suggests a picture in which the average number of satellite AGN 
increases with $M_h$ ($\alpha_s=0.22(-0.07,0.63)$ and $f_a^{\prime}=0.034(0.022,0.046)$) slower than the satellite
HOD of samples of galaxies ($\varpropto M_h^{\alpha_s=1-1.2}$).  
%with the satellites HOD of galaxies 
%slower, or may even increases, 
%than the satellite HOD of samples of galaxies
%$\varpropto M_h^{\alpha_s}$, with $\alpha_s \sim 1-1.2$.
%However models with $\varpropto M_h^{\alpha_s=1-1.2}$ 
%which describe the satellite HOD of galaxies, are not rejected ($\Delta \chi^2 < 2.3$).

%In addition, Fig.\ref{fig:HOD} compares the halo occupation of 41 XMM-AGN and
%41 XMM + 17 C-COSMOS AGN with the HOD model
%described in \citet{Miy11}. 
\citet{Miy11} first used a sample of ROSAT-RASS AGN and SDSS galaxies at 
$z \sim 0.3$ to study the occupancy of X-ray AGN in DM halos.
They investigate three models: the first (model a) assumes that AGN 
only reside in satellite galaxies while the others explore the 
effects of centrals (model b and c). In particular, in model b, 
the HOD of central AGN is constant and the satellite HOD has a 
power-law form at halo masses above $M_{min}$. Model c 
has the same form except that only less massive DM halos 
contain central AGN. Our results clearly show that only a 
model which includes AGN in both satellite and central galaxies 
at any halo mass $M > M_{min}$ can reproduce the observed AGN HOD.

In agreement with our results, they found that the upper limit of the power-law index 
of the satellite HOD is below unity, with $\alpha_s \leq 0.95$.
In detail, their model b with best fit parameters 
$\alpha_s=0.55$, log$M_{min}$[\hmsun] = 12.5 and
$f_a=2.9 \times 10^{-2}$ is in perfect agreement with our results,
showing that the luminosity and redshift evolution of the 
mean AGN HOD is not strong in the luminosity and redshift ranges of our interest.
In fact, we used an AGN sample with $\langle L_{X} \rangle = 10^{42.3}$ \ergs, while the
results in \citet{Miy11} provide the mean AGN HOD at $\langle z \rangle \sim 0.3$ for more
luminous AGN with $\langle L_{0.1-2.4} \rangle  \sim 10^{44.2}$ \ergs\
without any correction for the z evolution.
Then the two models suggest a similar positive $\alpha_s$ range,
but negative values of the slope are not rejected ($\Delta \chi^2 < 2.3$) in their model.
%They modelled the HOD of central AGN with a  
%constant and the satellite HOD with a power-law form at 
%halo masses above $M_{min}$ (model b). 
%Assuming $M_1/M_{min}$=23, they obtain $\alpha_s < 0.95$
%and $logM_{min} [h^{-1}M_{\odot}] = 12.5(12.2;13.0)$ 
%(68\% confidence interval is given in brackets).
Note that while in our case the AGN fraction is a free
parameter, they constrained it by normalizing the AGN HOD
to the observed AGN number density. 
%The two models
%suggest a similar $\alpha_s$ range, while the lower fraction
%of AGN might suggest a luminosity-dependent effect. \textbf{In fact,
%we used an AGN sample with $\langle L_{X} \rangle = 10^{42.3}$ $s^{-1}$erg, while the
%results in \citet{Miy11} provide the mean AGN HOD at $\langle z \rangle \sim 0.3$ for more
%luminous AGN with $\langle L_{0.1-2.4} \rangle  \sim 10^{44.2}$ $s^{-1}$erg
%without any correction for the z evolution.}
%Note that we fix $M_1=10^{14} M_{\odot}$, which is consistent
%with their choice $M_1/M_{min}$=23 motived by the HOD
%analysis of galaxies \citep{Zeh05}, with the best fit value of $M_{min}$
%equal to $10^{12.68} M_{\odot}$.

An assumption in the corrections applied by using Eq. \ref{eq:w} and \ref{eq:w0},
is that the shape of the XLF remains 
the same between AGN in groups and the general 
AGN population at $\log L_{\rm x}$[\ergs]$<42.4$. We still 
do not have sufficient observational basis to estimate the effects 
of possible difference in the XLF 
shapes between group environment and the field. At higher luminosities, 
Krumpe et al. (2010; 2012) 
found that AGN at log$\langle L_X \rangle$[\ergs] $\approx 44.6$ have higher 
bias values than those at $\approx 43.9$, corresponding to the mean 
host halo mass of $\log \langle M_{\rm h} \rangle$[\hmsun] = 
13.0 and 13.3 respectively \citep{Miy11}. Viewing this trend 
from the $M_{\rm h}$ dependence of the XLF shape, the XLF in high $M_{\rm h}$ 
environments have XLF more biased towards higher $L_{\rm X}$.
If the positive correlations between $L_{\rm X}$ and the host halo mass
$M_{\rm h}$ extends
to luminosities in the range $41.5<\log L_X$ [\ergs] $<42.4$, i.e. the range of 
$\log L_{\rm lim}(z)$ (see Fig. 1 left), the correction factors using the 
overall XLF in  Eq. 3 and 4 are overestimated
at higher $L_{\rm X}$ and underestimated at lower $L_{\rm X}$. A quantitative assessment of this effect
needs an estimate of the bivariate X-ray luminosity-host DMH mass function, which is far from being
available. We comment that, if the higher $L_X$ AGN preferentially occupy $\log M_{\rm h}$[\msun]$>13$ 
DMHs than the field, our estimate of $\alpha_{\rm s}$ should be corrected to a lower value.

\section{Two-points Statistics}\label{sec:ACFgroups}

\subsection{Method}\label{subsec:method}

The two-point auto-correlation function (ACF)
describes the excess probability over random of finding a pair 
with an object in the volume $dV_1$ and another in the volume $dV_2$, 
separated by a distance $r$ so that
$dP=n^2[1+\xi(r)]dV_1 dV_2$, where $n$ is the mean space density. 
A known effect when measuring pair separations is 
that the peculiar velocities combined with the Hubble flow
may cause a biased estimate of the distance
when using the spectroscopic redshift.  To avoid this effect
it is usually computed the projected ACF \citep{Dav83}: 
\begin{equation}\label{eq:wrp}
w_p(r_p)=2\int_{0}^{\pi_{max}}\xi(r_p,\pi)d\pi
\end{equation} 
where $r_p$ is the distance component perpendicular to the line of sight and 
$\pi$ parallel to the line of sight \citep{Fis94}.

The ACF is estimated by using the minimum variance estimator 
described by \citet{Lan93}:
\begin{equation}
 \xi(r_p,\pi)=\frac{DD-2DR+RR}{RR}
\label{eq:landy}
\end{equation}
where DD, DR and RR are the normalized  number of
data-data, data-random, and random-random source pairs, respectively. 
Eq. \ref{eq:landy}  indicates that an accurate estimate of the 
distribution function of the random  samples is crucial in order to 
obtain a reliable  estimate of $\xi(r_p,\pi)$. 
%Note that other estimator have been proposed in the literature, but
%the  \citet{Lan93} one has been shown to provide the smallest statistical variance. 
%Such a formalism can be easily adopted when computing the angular or the redshift 
%space correlation function, with the only difference that the evaluation is made on  a single 
%dimension. 
Several observational biases must be taken 
into account when generating a random sample of AGN in a X-ray flux limited survey. 
In particular, in order to reproduce the selection function of the survey,
one has to carefully reproduce the space and flux distributions 
of AGN, since the sensitivity in X-ray surveys is not 
homogeneous on the detector and therefore on the sky.
%This points out the necessity of create a random sample which includes as many selection effects
%as possible since the estimate of $\xi(r)$ (or $w(\theta)$) is strongly dependent on RR (see eq. 
%\ref{eq:landy}. Moreover in several case optical follow-up of the X-ray source is not 100\% complete,
%therefore one must carefully reproduce the mask effect. What is usually done is that to 
%create random samples in 3D, sources are placed at the same angular position of the 
%real sources and redshift are randomly drawn from a smoothed redshift distribution 
%of the real sources.  If instead the spectral completeness is close to 100\% then 
%the right procedures is to occupy the survey volume with  random sources drawn 
%from a L-z dependent luminosity function and accept check if they would be 
%observable using a sensitivity map.
An other important choice for obtaining a reliable estimate of $w_p(r_{p})$,
is to set $\pi_{max}$ in the calculation of the integral above. 
One should avoid values of  $\pi_{max}$ too large since they would  add noise to
the estimate of $w_p(r_p)$.  If, instead, $\pi_{max}$ is too small  one could not recover
all the signal. 
%Uncertainties in the ACF are usually evaluated with a bootstrap resampling technique
%but  it is worth noting that in the literature, several methods  
%are adopted for errors estimates in two-point statistics \citet[for a detailed description]{Nor09}.
%It is  known  that Poisson estimators generally underestimate the variance because 
%they do consider that points in ACF are not  statistically independent. 
%Jackknife resampling method, where one divides the survey area in many
%sub fields and iteratively re-computes correlation functions by
%excluding one sub-field at a time, generally gives a good estimates of 
%errors. But it requires that sufficient number of almost statistically
%independent sub- fields, this is not the case  for most of X-ray surveys where the
%source statistics is moderately low. 
%\citet{Coi09} estimated the error bars on the two-point correlation function including both Poisson 
%and cosmic variance errors estimated, using DEEP2 mock catalogs derived from the Millenium Run simulations.

We created an AGN random sample where each simulated source is placed at 
a random position in the sky, with flux randomly extracted from the catalog 
of real source fluxes. The simulated source is kept in the random sample 
if its flux is above the sensitivity map value at that position \citep{Miy07, Cap09}.
The corresponding redshift for a random object is assigned based 
on the smoothed redshift distribution of the AGN sample.

Similarly, an unclustered catalog of galaxy groups mimicking 
the selection function of the survey must be employed to quantify the degree 
to which the groups preferentially locate themselves in one another's neighborhood.
%Using the weak lensing calibration of the sample and accounting for the
%sensitivity map towards the search of extended X-ray emission, we were able
%to achieve a good match between the observed $dN/dz$ relation and the
%prediction of the WMAP7 $\Lambda$CDM model \citep[][Finoguenov et al. in prep]{Fin10; Bie10}.
%Based on the success of the model,
We generate a random group catalog by calculating at each area of a given
sensitivity, the probability of observing a group of a given mass and redshift.
We then used Monte-Carlo simulation of the group positional assignment,
finally producing a catalog of hundred thousand objects. Also the X-ray surface
brightness sensitivity map is non-uniform in depth and consequently the
probability of detecting groups of a particular mass is variable with
redshift; in particular the minimum mass below which a group will be
detected is an increasing function of $z$.

\subsection{Galaxy Groups ACF}

\begin{figure}
\plotone{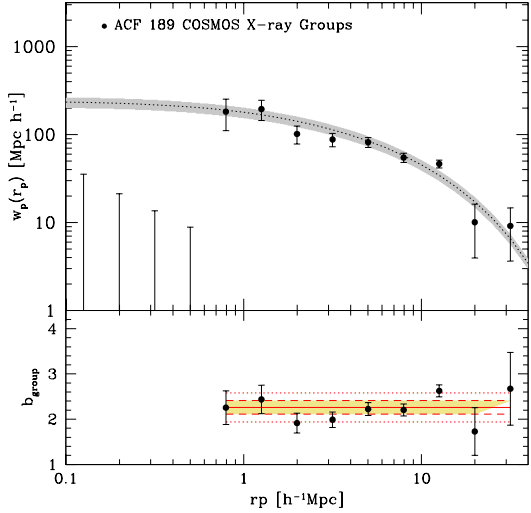}
\caption{\footnotesize :\textit{Upper Panel}: Projected auto-correlation function of galaxy groups with $z \leq 1$. Following the halo model approach, the clustering signal at large scale ($r_p > 1-2h^{-1}$ Mpc) is produced by separate DM halos (so called 2-halo term) and it can be expressed as $b_{group}^2w_{DM}^{2-h}(r_p,z=0)$ (dotted line), where the bias factor is defined in Equation \ref{Eq:bg}. \textit{Lower Panel}: Bias factor versus $r_p$. The solid line shows the best-fitting constant value. The shaded regions indicate the values of the bias for which $\Delta \chi^2=1$ (dashed line) and $\Delta \chi^2$ = 4 (dotted lines).}
\label{fig:acfgroups}
\end{figure}

We measured the projected ACF of galaxy groups
in the range $r_p=0.1-40$ \mpch\ by using Eq. \ref{eq:wrp},
with $\pi_{max}=80$ \mpch\ (see Fig.\ref{fig:acfgroups}). 
The errors have been estimated using bootstrap resampling of the data, 
which consists of computing the variance of $w_p(r_p)$ in $N_{real}$
bootstrap realizations of the sample.
Each realization is obtained by randomly selecting a subset of 
groups from the data sample allowing for repetitions.

In the halo model approach, 
the clustering signal can be
modelled as the sum of two contributions of pairs from the
same DM halo (1-halo term) and those from different DM halos (2-halo term).
In Fourier space, the 2-halo term can be explicitly written as (Seljak 2000, Cooray \& Sheth 2002):
\begin{equation}
P_{2-h} \approx b^2 P_{lin} (k,z)
\end{equation}
where $P_{lin}(k,z)$ is the linear power spectrum and $b$ is the bias
factor of the sample.
Then the galaxy groups two-point correlation function at large scales is
given by:
\begin{equation}\label{eq:b1}
%w_{DM}^{2-h} (r_p) = \int \frac{k}{2 \pi} P_{lin} (k) J_{0}(kr_p)dk
w_{p,2-h} (r_p) = b_{group}^2 \int \frac{k}{2 \pi} P_{lin} (k) J_{0}(kr_p)dk
\end{equation}
where $J_{0}(x)$ is the zeroth-order Bessel function of the first kind.
Following this model, the galaxy group bias defines the relation between
the 2-halo term of DM and groups clustering signal: 
\begin{equation}\label{eq:b}
b^2_{group,obs} (r_p)=\frac{w_{p,2-h}(r_p)}{w_{DM}^{2-h}(r_p,z=0)}
\end{equation}
According to this equation, we estimated an average bias factor 
in the range $r_p=1-40$ \mpch\ equal to
$b_{group,obs}=2.20 \pm 0.12$ where the error corresponds
to $\Delta \chi^2=1$ using a $\chi^2$ minimization technique with
1 free parameter. Following the bias mass relation $b(M_{h},z)$ described in 
\citet{van02} and \citet{She01}, this observed bias, with respect 
to the DM distribution at $z=0$ corresponds to a typical mass
log$M_{typ}$ [\hmsun] = 13.65$^{+0.07}_{-0.08}$.

On the other hand the average bias of galaxy groups 
can be estimated starting
from the known masses $M_{200}$.
Since usually the bias factor is related to the average
DM halo mass expressed in units of $h^{-1}M_{\odot}$,
hereafter we refer to $M_h$ to indicate the galaxy groups masses
in these units.
%Following the halo model, the large scale bias of
%galaxy groups at redshift $z$ is given by:
%\begin{equation}\label{eq:biasg}
%b_{group} (z)= \frac{ \int_{M_h} b_h(M_h,z)\langle N \rangle(M_h)n_h(M_h,z)dM_h} {\int_{M_h}\langle N \rangle(M_h)n_h(M_h,z)dM_h}
%\end{equation}
%Since the HOD of galaxy groups is by definition a delta function
%$\delta(M-M_{h})$, Eq. \ref{eq:biasg} becomes
%\begin{equation}\label{eq:bg}
%b_{group}(z)=b_h(M_{h},z)
%\end{equation}
%where $b(M_{h},z_i)$ is evaluated using \citet{van02} and \citet{She01}
%and $M_{h}$ are in units of $h^{-1}M_{\odot}$. \\
Since we know the mass estimates, we can predict the 
bias factor of this sample of galaxy groups. 
By accounting for the fact that the linear regime of the structure
formation is verified only at large scales,
we estimated the average bias of the sample, including only
pairs which contribute to the clustering signal at $r_p=1-40$ Mpc $h^{-1}$.
As described in \citet{Alle11}, we measured the average bias factor of the sample as:
\begin{equation}\label{Eq:bg}
b_{group}(M_{h}) = \sqrt{\frac{\sum_{i,j}b_{group,i}b_{group,j} D_iD_j } {N_{pairs}} }
\end{equation}
where $b_{gg,i}b_{gg,j}$ is the bias factor of the group pair $i-j$ and $N_{pairs}$
is the total number of group pairs in the range $r_p=1-40$ Mpc $h^{-1}$.
The $D$ factor is defined by $D_1(z)/D_1(z=0)$,
where $D_1(z)$ is the growth function (see eq. (10) in \citet{Eis99} and references therein) and
takes into account that the amplitude of the DM 2-halo term
decreases with increasing redshift.\\
%Similarly, we defined an average mass of galaxy groups
%weighting the mass of each pair for the $D$ factor and
%the bias of the pair ($b_{group,i}b_{group,j}$):
%\begin{equation}
%\langle M_h \rangle = \frac{\sum_{i,j}b_{group,i}b_{group,j} D_iD_j M_h^i M_h^j} {\sum_{i,j}b_{group,i}b_{group,j}D_iD_j}
%\end{equation}
%where $M_h$ is the galaxy group mass in units of $h^{-1}M_{\odot}$.
By using this approach we obtained $b_{group}=2.21^{+0.13}_{-0.14}$ 
where the errors have been estimated assuming a $20\%$ error on the 
galaxy groups masses. This value is in perfect agreement with the bias obtained from the ACF,
with $b_{group}/b_{group,obs} = 1.00 \pm 0.05$. 
%and $\langle M \rangle =13.61^{+0.09}_{-0.10}h^{-1}M_{\odot}$, 

\subsection{X-ray AGN-Galaxy Groups Cross-Correlation}\label{sec:CCF}

Measurements of the cross-correlation function between AGN and groups use a
version of the estimator proposed by \citet{Lan93}:
\begin{equation}\label{eq:LZ2}
\xi = \frac{1}{R_g R_A} (D_gD_A - D_g R_A- D_A R_g + R_g R_A)
\end{equation}
where each data sample, with pair counts $D_i$ has an associated 
random catalog, with pair counts $R_i$ normalized
by its number density.
%An accurate estimate of the distribution function of the random samples
%is crucial and several observational biases must be taken into account when 
%generating a random sample of objects in a flux limited survey.
%In particular, in order to reproduce the selection function of the survey, one has to 
%carefully reproduce the space and flux distributions of the sources, since the sensitivity 
%of the survey are not usually homogeneous on the sky. Simulated AGN must 
%be randomly placed on the survey area, and the procedure must be 
%repeated hundreds times per sample in order to have a significant 
%sampling of random realizations to estimate the errors.

%A known effect when measuring pairs separations is that the peculiar 
%velocities combined with the Hubble flow may cause a biased estimate 
%of the distance when using the spectroscopic redshift. To avoid this effect 
%it is usually computed the projected ACF \citep{Dav83}: 
%\begin{equation}\label{eq:wrp}
%w(r_p) = 2\int_0^{\pi_{max}} \xi(r_p,\pi)d\pi
%\end{equation}
%where $r_p$ is the distance component perpendicular to the line of sight 
%and $\pi$ parallel to the line of sight \citep{Fis94} . \\

%\section{Large-scale AGN Bias in HOD Model}\label{subsec:AGNbias}

\begin{figure}
\plotone{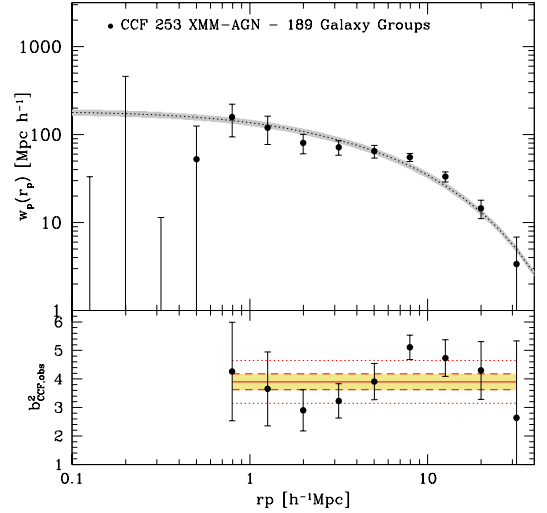}
\caption{\footnotesize : \textit{Upper Panel}: Projected cross-correlation function of 253 XMM-COSMOS AGN with $z_{spec} \leq 1$ and 189 galaxy groups, excluding from the analysis AGN that are within in galaxy groups. The clustering signal at large scale $r_p> 1-2h^{-1}$ Mpc is due to AGN residing in different DM halos and is described by $b_{CCF}^2w_{DM}^{2-h}(r_p,z=0)$ (dotted line), where the bias factor is defined in Equation \ref{eq:biasccf}. On the contrary the 1-halo term is zero since we are removing from the analysis AGN with $R_{200}$ responsible for the signal of AGN within the same DM halos. \textit{Lower Panel}: Bias factor versus $r_p$. The solid line shows the best-fitting constant value. The shaded regions indicate the values of the bias for which $\Delta \chi^2=1$ (dashed line) and $\Delta \chi^2$ = 4 (dotted lines).}
\label{fig:ccf}
\end{figure}

We estimated the cross-correlation function of 189 galaxy groups 
and a subset of 253 XMM-COSMOS AGN with known spectroscopic redshift $\leq 1$, 
obtained excluding those within $R_{200}$ (see Fig. \ref{fig:ccf}).
%In the linear regime,
%the clustering signal can be
%modelled as the sum of two contributions of pairs from the
%same DM halo (1-halo term) and those from different DM halos (2-halo term).
Following Eq. \ref{eq:b1}, the linear bias factor of the projected CCF of AGN with galaxy groups
can be approximated by using the 2-halo term:
\begin{equation}\label{eq:biasccf}
b^2_{CCF,obs} (r_p)=b_{AGN}(r_p) b_{group}(r_p) = \frac{w_{p,2-h}^{CCF}(r_p)}{w_{DM}(r_p,z=0)}
\end{equation}
where $b_{AGN}$ and $b_{group}$ are the bias factor of AGN and
galaxy groups, respectively and $w_{DM}$ is the projected dark matter CF.
%For the linear power-spectrum, $P_{lin}(k,z)$ we use the primordial
%power spectrum with $n_s=1$ and a transfer function calculated using \cite{Eis98}.}
%can be estimated using the clustering signal
%at large scales:
%\begin{equation}\label{eq:biasccf}
%b^2_{CCF,obs} (r_p)=b_{AGN}(r_p) \times b_{group}(r_p) = \frac{w_{CCF}(r_p)}{w_{DM}(r_p,z=0)}
%\end{equation}
%where $b_{AGN}$ and $b_{group}$ are the bias factor of AGN and
%galaxy groups, respectively and $w_{DM}$ is the projected dark matter CF.
Fig. \ref{fig:ccf} (\textit{Lower Panel}) shows the linear bias $b_{CCF}^2$ as a function of $r_p$
over the scales $r_p \sim 1-40$ \mpch. We fitted the data points with a 
constant by using the $\chi^2$ minimization technique 
and we found $b^2_{CCF,obs}=3.90 \pm 0.28$. 
The shaded regions show the bias values for which
$\Delta \chi^2$=1 and 4
(68\% and 99\% confidence levels for one parameter).

\section{The bias factor in the HOD model}\label{subsec:}

Based on the halo model approach, the AGN bias factor
depends on the mean AGN HOD $\langle N_{AGN} \rangle(M_h)$:
\begin{equation}\label{eq:bHOD}
b_{AGN}(z)= \frac{\int_{M_{min}}^{\infty}b_h(M_h,z)\langle N_{AGN} \rangle(M_h)n_h(M_h,z)dM_h} {\int_{M_{min}}^{\infty}\langle N_{AGN} \rangle(M_h)n_h(M_h,z)dM_h}
\end{equation}
where $M_{min}$ is the minimum mass below which the AGN HOD is zero,
$n_h(M_h)$ and $b_h(M_h)$ are the halo mass function and the halo bias 
given by \citet{She01}.
Note that the large-scale bias factor does not depend on 
the normalization log$M_1 [M_{\odot}] = 10^{13.8}$ and on $f_{a}$.
Since we are estimating the CCF 
excluding from the XMM-COSMOS sample 41 AGN in galaxy groups,
the clustering signal at large scale is due to AGN that live in halos
with $M_{200}<10^{13}M_{\odot}$ or with masses that
we can not observe at a given redshift $z$. 
This implies that the AGN bias can be written as:
\begin{equation}\label{eq:bAGN}
b_{AGN}(z)= \frac{\int_{M_{min}}^{M_x(z)}b_h(M_h,z)\langle N_{AGN} \rangle(M_h)n_h(M_h,z)dM_h} {\int_{M_{min}}^{M_x(z)}\langle N_{AGN} \rangle(M_h)n_h(M_h,z)dM_h}
\end{equation}
where $M_x(z)$ is the minimum mass we can observe for a group at redshift $z$
(at $z \sim 1$ we only detect luminous and then massive groups).
Note that we are assuming the separability of the mass and redshift dependence 
of the $\langle N_{AGN} \rangle$, i.e.
$\langle N_{AGN} \rangle (M_{h},z)=\langle N_{AGN} \rangle (M_{h},z=0) \times \langle N_{AGN} \rangle(z)$. 
%and we took into account the $1\sigma$ error on the power-law
%index $\alpha$ and $b_{group}$.

%We found can then constrain the $M_{min}$ value that reproduces
%the observed bias $b^2_{CCF,obs}$ estimated with Eq. \ref{eq:biasccf}, i.e. such that:
Following this method, the bias factor $b^2_{CCF}$ is defined by:
\begin{equation}
b^2_{CCF}=\frac{\sum_{i,j}b_{AGN,i}b_{group,j}D_iD_j}{N_{pairs}}
\end{equation}
where the sum is over the pairs $i,j$ contributing to the 
clustering signal at large scale and $b_{AGN}(z_i)$ is the AGN bias 
(Eq. \ref{eq:bAGN}) assuming a rolling-off power-law 
HOD based on our results. $b_{group,j}$ is the bias associated
to the galaxy group mass and redshift following \citet{She01} and
$N_{pairs}$ is the total number of AGN-group pairs 
in the range $r_p=1-40$ \mpch.
We found $b^2_{CCF}=3.97^{+0.10}_{-0.05}$, which is in
perfect agreement with the observed bias factor defined in Eq. \ref{eq:biasccf}. 
The errors are due to the $1\sigma$ errors
on the power-law index $\alpha$, $M_{min}$ and on $b_{group}$.
%$b_{ccf,obs}^2/b_{ccf}^2=$. 

%\begin{figure}
%\plotone{plots/fig8.eps}
%\caption{\footnotesize : \textit{Upper Panel}: Mean AGN occupation as a function of halo mass, by
%using the best fit (solid line), the lowest (dotted line) and highest (dashed line) value
%of $\alpha$ in the $1\sigma$ confidence interval. \textit{Lower Panel}: 
%$\alpha-M_{min}$ relation corresponding to the best-fitting value of $\alpha$, $b_{group}$ and $b^2_{CCF,obs}$ 
%(black crossed) and when including the 1$\sigma$ errors on these parameters
%(shaded region). The red lines indicate the combination of $\alpha-M_{min}$ values 
%along a constant average mass $\langle M \rangle(z=0)$ defined in Eq. \ref{eq:aveM}.}
%\label{fig:HODlast}
%\end{figure}
%We found a minimum mass above which a DM halo hosts an AGN
%equal to $logM_{min} [h^{-1}M_{\odot}] = 12.20 (11.65; 12.45) $, 
%where the errors are due to the $1\sigma$ error on the power-law HOD index
%and on $b_{group}$ and $b^2_{CCF}$.
%Fig. \ref{fig:HODlast} shows the AGN HOD 
%as function of halo mass when using the best fit (solid line),
%the lowest (dotted line) and highest (dashed line) value of $\alpha$.
%The smaller is the power-law index, the higher is
%the minimum mass below which the AGN HOD is zero. 
%The black cross in the \textit{Lower Panel} of Fig. \ref{fig:HODlast} indicates the $\alpha-M_{min}$ 
%combination corresponding to the best-fitting constant $b^2_{CCF}$,
%while the shaded region shows the $\alpha-M_{min}$ 
%relation when we include the 1$\sigma$ error on $\alpha$, $b_{group}$
%and $b^2_{CCF}$. \\
Similarly we can define the average halo mass $\langle M \rangle$
corresponding to the observed CCF signal, i.e.:
\begin{equation}\label{eq:aveM}
\langle M \rangle = \frac{\sum_{i,j}b_{AGN,i}b_{group,j}D_iD_jM_h^{i,j}} {\sum_{i,j}b_{AGN,i}b_{group,j}D_iD_j}
\end{equation}
%where the sum is over the pairs $i,j$ contributing to the 
%clustering signal at large scale, $b_{AGN,i}$ is defined by 
%Eq. \ref{eq:bAGN}, $b_{group,j}$ is the bias associated
%to the galaxy group mass and redshift, following
%\citet{van02} and \citet{She01}.
where $M_h^i (z_i)$ is the average AGN host halo mass at redshift $z_i$:
\begin{equation}
M_h^i (z_i)= \frac{\int_{M_{min}}^{M_x(z)} \langle N_{AGN} \rangle(M_h)n_h(M_h,z_i)M_h dM_h} {\int_{M_{min}}^{M_x(z)}\langle N_{AGN} \rangle n_h(M_h,z_i)dM_h}
\end{equation}
while $M_h^j(z_j)$ is the galaxy group mass, both in units of \hmsun.
Fixing the values of $\alpha$ and $M_{min}$, we found 
the average mass of XMM-COSMOS AGN to be 
$log \langle M \rangle$ [\hmsun]= 13.20(13.10;13.25), where the errors
correspond to the 68\% confidence region in the log$M_{min}-\alpha$ space.
%shown in fig. \ref{fig:HODlast} (\textit{Lower Panel}). 
%Despite the large errors on $\alpha$ and $M_{min}$,
%$\langle M \rangle$ is determined with a small uncertainty of $\sim$0.1 dex,
%since the confidence region roughly follows a constant
%halo mass track.

\section{Discussion}\label{sec:disc}

Using a sample of X-ray selected AGN and galaxy groups
in the COSMOS field at $z \leq 1$, we performed the first direct 
measurement of the AGN HOD in the mass range 
$logM_{200}=13-14.5M_{\odot}$, based on the mass function
of galaxy groups hosting AGN. In contrast to previous works using
the clustering signal of the sample, we directly counted the 
number of AGN within galaxy groups and we found 58 AGN
in groups, associated to 22 central and 36 satellite galaxies. 
This allowed us to put constrains on both the mean occupation of AGN among satellite
and central galaxies as function of the halo mass,
which provides information on the AGN triggering mechanism.
\citet{Sta11} studied the halo occupation properties of  
AGN detected by the Chandra X-ray Observatory in the Bootes field over a redshift interval from z=0.17 to z=3,
showing that X-ray AGNs are predominantly located at the centers of DMHs with 
$M_h > 4.1 \times 10^{12}$ \hmsun, with an upper limit
of the satellite fraction of 0.1 ($\Delta \chi^2<2.3$).
The central locations of the quasar host galaxies are expected in major merger 
models because mergers of equally sized galaxies preferentially occur at the 
centers of DMHs (Hopkins et al. 2008).
On the contrary \citet{Pad09} observed the presence 
of the one-halo term in the cross-correlation function of optically selected 
$z < 0.6$ quasars and Luminous Red Galaxies, and use this to conclude 
that a large fraction of the AGNs is hosted by satellite galaxies.

Our results show that the average number of AGN in satellite galaxies
in the halo mass range $log M_h [M_{\odot}]=13-14.5$,
and AGN luminosity $log \langle L_X \rangle$ [\ergs]= 42.3
might be comparable or even larger that the average number of 
AGN in central galaxies, i.e. X-ray AGN do not avoid satellite galaxies.
A high fraction of AGN in satellite galaxies is expected in a
picture where other phenomena like secular processes,
might become dominant in the AGN activation. 
\citet{Mil06}, \citet{Hop06} and \citet{Hop09} showed that
low-luminosity AGN could be triggered in more common nonmerger events, 
like stochastic encounters of the black holes and molecular clouds, 
tidal disruption or disk instability. This leads to the expectation of a 
characteristic transition to merger-induced fueling around the traditional 
quasar-Seyfert luminosity division.

Moreover we found the power-law slope, which defines the evolution of the
mean satellite HOD with halo mass, to be $\alpha_s \sim 0-0.6$,
suggesting a picture in which the 
average number of satellite AGN per halo increases with the halo mass. 
On the other hand, \citet{Miy11} obtained $\alpha_s < 0.95$,
but negative values of the slope are not rejected ($\Delta \chi^2 < 2.3$).

It is interesting to compare this result with HOD analyses of galaxies.
Previous HOD analyses of galaxies found $\alpha_s \sim 1-1.2$ 
for a wide range of absolute magnitudes and redshifts at least up to 
$z \sim 1.2$ \citep{Zeh05, Zhe07, Zeh10}, implying a simple
proportionality between halo mass and satellite number, $\langle N_{sat}\rangle \propto M_h$.
Our results suggest that the mean HOD of satellite AGN might
increase slower ($\alpha_s < 0.63$) with the halo mass respect to 
the linear proportion ($\alpha_s = 1$) in the satellite galaxy HOD,  i.e. the AGN
is not only triggered by the halo mass.
On the contrary, a decreasing AGN fraction with the halo mass is 
consistent with previous observations that the 
AGN fraction is smaller in clusters than in groups in the 
nearby universe.

In order to fully understand the growth history of SMBHs as
well as the physical processes responsible for the AGN activity
we need to explore the AGN HOD at
different redshifts, luminosities, and AGN types.
The luminosity distribution of AGN that reside 
in halos of a given mass provides a tool to examine the distribution
of halo mass for a given luminosity and study luminosity dependent
clustering. While this formalism has been widely used in modelling
galaxy clustering, it is still not applicable to AGN. In fact,
it is also important to have larger numbers of AGN in galaxy groups
which will enable stronger constraints on the shape of the satellite 
and central AGN HOD, hopefully as function of AGN luminosity and redshift.

\section{Conclusions}\label{sec:concl}

We have performed the first direct measurement of the mean
halo occupation distribution of X-ray AGN as function of halo mass,
by directly counting the number of AGN within X-ray galaxy groups
with masses $logM_{200}[M_{\odot}] = 13- 14.5$, in the COSMOS field at $z \leq 1$.
Our findings are summarized as follows.
\begin{enumerate}
  \item We identified 41 XMM-COSMOS AGN within 
  galaxy groups, defined as AGN located within 3 times the group
  line-of-sight velocity dispersion and within $R_{200}$ and 17
  additional sources including in the analysis C-COSMOS only selected AGN.
   \item We measured the mean AGN occupancy of galaxy groups 
   as function of halo mass in the range log$M_{200}[M_{\odot}] = 13- 14.5$ 
   and we modelled the data points with a rolling-off power-law with the best fit index
   $\alpha=0.06(−0.22, 0.36)$ and normalization parameter $f_a=0.05(0.04, 0.06)$.
 \item Using a galaxy membership catalog, we associated 22/58 and 36/58
 AGN to central and satellites galaxies, respectively. 
 We constrained that the mean AGN occupation function among central galaxies 
 is described by a softened step function above log$M_{min}=12.75(12.10, 12.95)$
 while the satellite AGN HOD increases with the halo mass
 ($\alpha_s<0.63$) slower than the satellite HOD of sample of
 galaxies ($\alpha_s=1-1.2$).
  \item  We presented an estimate of the projected ACF
  of galaxy groups over the range $r_p =$ 0.1-40 \mpch\
  at $\langle z \rangle=0.5$. 
  We verified that the bias factor and the corresponding typical halo mass 
  estimated from the observed galaxy group ACF,
  are in perfect agreement with the values $b_{group}$ and $\langle M_h \rangle$ obtained
  by using the galaxy group mass estimates.
  In particular we found $b_{group}=2.21^{+0.13}_{-0.14}$ and log $\langle M_h \rangle$ [\hmsun] = 13.61$^{+0.09}_{-0.10}$,
%  \item We performed the first measurement of the projected AGN-galaxy groups cross-correlation function,
%  excluding from the analysis AGN that are within $R_{200}$. Interpreting the large-scale
%  clustering signal with the halo model approach, we estimated $b_{CCF}^2=3.9 \pm 0.3$ and 
%  fixing this value among with the AGN HOD,
%  we constrained that the minimum mass above which a halo hosts an AGN is 
%  $logM_{min} [h^{-1}M_{\odot}] = 12.20(11.65; 12.45) $.
%  \item Based on our results of the AGN HOD, we constrained the average
%  mass of the hosting halos for XMM-COSMOS AGN that do not reside in galaxy groups to
%  log$\langle M \rangle [h^{-1}M_{\odot}] = 12.90^{+0.15}_{-0.12}$.
\end{enumerate}

\newpage

\begin{deluxetable*}{llllllllllll}
\tabletypesize{\scriptsize}
\tablewidth{0pt}
\tablecaption{AGN in galaxy groups \label{tbl-2}}
\tablehead{
\colhead{$ID$} &
\colhead{Ra} &
\colhead{Dec} &
\colhead{$z$}&
\colhead{log$M_{200}$\tablenotemark{a}} &
\colhead{$ID$} &
\colhead{AGN Ra} & 
\colhead{AGN Dec} &
\colhead{$ID$} &
\colhead{$z$\tablenotemark{c}} &
\colhead{$log L_X$} &
\colhead{Flag} \\
\colhead{Groups} &
\colhead{deg} &
\colhead{deg} &
\colhead{Groups}&
\colhead{[$M_{\odot}$]} &
\colhead{$XMM$\tablenotemark{b}} &
\colhead{deg} & 
\colhead{deg} &
\colhead{$Chandra$\tablenotemark{b}} &
\colhead{AGN} &
\colhead{[$h^{-2}_{70}$ erg $s^{-1}$]} &
\colhead{\tablenotemark{d}} }
\startdata
\multicolumn{12}{l}{}\\
  11  &  150.1898   & 1.65725& 0.22  &  14.28   & -99    & 150.19864  & 1.671924  & 12390& 0.228$^*$  & 41.27   & 0   \\
  17  &  149.96413 & 1.68033& 0.372 & 14.02   & 30744& 149.96396  & 1.6805983& -99    & 0.372  &    43.28 & 1   \\
  19  &  150.37282 & 1.60944&  0.103 &12.98   &  2021  & 150.37256 & 1.6093965 & 1678  & 0.104  &   41.22 &  1   \\
  20 &   150.32494 & 1.60313   & 0.22  & 13.38 & 2186  &150.33601 & 1.6012201 & 1671  & 0.234$^*$     & 41.88 & 0   \\
  35 &   150.2066  & 1.82327   & 0.53 & 13.70 & -99  & 150.20702 & 1.823398  & 1292   & 0.529 & 41.73    &          1   \\
  39 &   149.82381 & 1.8252701 & 0.531& 13.63 & 5502 & 149.81248 & 1.8238187 & 219 & 0.529 & 42.37 & 0   \\
  52 &   150.44704 & 1.8828501 & 0.671   &   13.64 & 30681 & 150.44731 & 1.8832886 & 690  & 0.670$^*$ & 42.75 & 1   \\
  69   &  150.42012 & 1.9708 & 0.862 &  13.49 & 5519   & 150.42473 & 1.9693011 & 1185 & 0.854$^*$ &  42.77 & 0 \\
  78 &  150.27461   & 1.98884 & 0.838   & 13.61 &  -99 & 150.27823 & 1.990302  & 685  & 0.838 & 42.35    &          1   \\
  78   &  150.27461 & 1.9888 & 0.838 &  13.61 & 206     & 150.27434 & 1.9884306 & 1100 & 0.673 &  42.57 & 0 \\
  79 &   150.44728 & 2.05392   & 0.323 & 13.77 & -99 &  150.44765 & 2.053961  & 2605   & 0.323  &    41.62     &         1   \\
  87 &   150.51109 & 2.0269899 & 0.899 & 13.60 &  2387  & 150.51036 & 2.0293686 & 496  &   0.899  &    43.15    &          1   \\
  93   &  149.6692   & 2.0740 & 0.338 &  13.59 & 417     & 149.6692   & 2.073962   & 168   & 0.34   &  42.76 & 0 \\
  110 &  150.17979 & 2.1103699 & 0.361    &  13.43 &   6   &  150.17978 & 2.1101542 & 42   &   0.360 & 43.19 &  1   \\
  118 &  149.63463 & 2.1357 & 0.962 &  13.70 & 411     & 149.63828 & 2.1494889 & 323   & 0.952 &  43.25 & 0 \\
  124 &  150.05656 & 2.2085 & 0.186 &  13.22 & 302     & 150.0571   & 2.2063098 & 1297 & 0.186 &  41.25 & 0 \\
  124 &  150.05656 & 2.20854  &  0.186  & 13.22 & -99 & 150.09554 & 2.2202671 & 1221 & 0.186  &    40.86     &         0   \\
  128 &  150.58435 & 2.1811 & 0.556 &  13.58 & 2855   & 150.57634 & 2.1812408 & 1634 & 0.554 &  42.41 & 0 \\  
  127 &  150.44104 & 2.15873 &  0.377   &   13.32 & -99  & 150.44202 & 2.1551671 & 2313  & 0.376  &   41.24    &          1    \\          
  130 &  150.02382 & 2.2032 & 0.942 &  13.84 & -99     & 150.02303 & 2.206567   & 2413 & 0.94   &  42.45 & 0 \\         
  136 &  150.17493 & 2.2170 & 0.676 &  13.56 & -99     & 150.17592 & 2.2156539 & 877   & 0.667$^*$ &  42.08 & 0 \\ 
  137 &  149.96271 & 2.2102399 & 0.425 & 13.32&  -99  & 149.96216 & 2.2102211 & 12011 & 0.425 &     41.44     &         1   \\
  138 &  149.50874 & 2.2614 &   0.943 & 13.97 & 5539 &  149.50914 & 2.261276 &  -99  &  0.962 & 43.19      &        1   \\
  142 &  150.28798 & 2.2769 & 0.122 &  13.09 & 317     & 150.25331 & 2.2779887 & 3718 & 0.165 &  41.08 & 0 \\            
  143 &  150.21454 & 2.2801 & 0.880 &  13.77 & 116     & 150.20604 & 2.2857771 & 22     & 0.874 &  43.33 & 0 \\
  149 &  150.41566 & 2.4302 & 0.124 &  13.85 & 45       & 150.33598 & 2.4335926 & 658   & 0.121 &  41.31 & 0 \\ 
  149 &  150.41566 & 2.4302001 & 0.124   &   13.85 & 73  &  150.416 &  2.4299896 & 634 &   0.124 &     42.00 &  1 \\   
  161 &  149.95262 & 2.3418 & 0.941 &  13.74 & 430     & 149.95949 & 2.3560882 & 497   & 0.889 &  42.85 & 0 \\
  171 &  149.66328 & 2.2677 & 0.676 &  13.56 & 5411   & 149.66243 & 2.2693584 & 1646 & 0.676 &  42.76 & 0\\         
  173 &  150.05804 & 2.38045 &  0.347 & 13.48 & 241 &  150.05794 & 2.3805208 & 898  &  0.347  &    42.14 &  1  \\  
  174 &  149.63988 & 2.3491 & 0.950 &  13.97 & 53937 & 149.64459 & 2.3587193 & -99   & 0.962$^*$ &  42.88 & 0 \\ 
  175 &  150.24123 &2.34835 &  0.723 &13.55 & -99 &  150.24109 &2.3483839 &977  &  0.703 & 41.72   &           0   \\
  186 &  150.21748 & 2.4003 & 0.905 &  13.77 & 53303 & 150.21141 & 2.4022212 & 975   & 0.894$^*$ &  42.62 & 0 \\
  194 &  149.69957 & 2.4028 & 0.354 &  13.45 & 135     & 149.70084 & 2.4025679 & 417   & 0.376 &  43.00 & 0 \\ 
  196 &  150.27898 & 2.4192 & 0.123 &  13.11 &  -99    & 150.27975 & 2.4222209 & 1315 & 0.122 &  40.25 & 0 \\        
  216 &  150.06664 & 2.6474 & 0.696 &  13.64 & 5113   & 150.06633 & 2.642817   & 143   & 0.693 &  42.45 & 0 \\ 
  217 &  150.00713 & 2.4534299 & 0.731 & 13.54 & -99 &  150.00188 & 2.4606521 & 1243  & 0.732 & 42.39   &           1   \\
  219 &  150.27148 & 2.5134399 & 0.704  &    13.54 & 158 &  150.27411 & 2.5117824 & 138 &   0.703 & 42.51 &  1 \\  
  220 &  149.92343 & 2.5249 & 0.729 &  14.39 & 486     & 149.92007 & 2.5143571 & 562   & 0.698 &  42.52 & 0 \\
  220 &  149.92343 & 2.5249901 & 0.729 & 14.39 & -99  & 149.91588 & 2.52139 &   1310 &  0.729 & 42.38      &        0   \\
  231 &  150.05421 & 2.58885 &  0.675 & 13.73 &  8 &    150.05383 & 2.5896702 & 142 &   0.699  &    44.03 &   1   \\
  234 &  150.15816 & 2.6082 & 0.893 &  13.66 & -99     & 150.15796 & 2.6113441 & 727   & 0.863$^*$ &  42.35 & 0 \\          
  237 &  150.11774 & 2.6842501 & 0.349 & 14.00 &  5118 & 150.11783 & 2.6840661 & -99  &  0.349 & 42.18 &  1   \\
  259 &  149.65717 & 2.8195 & 0.703 &  13.73 & 60270 & 149.65953 & 2.8270493 & -99   & 0.708$^*$ &  42.33 & 0 \\ 
  262 &  149.60007 & 2.8211 & 0.344 &  14.03 & 5320   & 149.63142 & 2.8174951 & -99   & 0.34   &  42.93 & 0 \\
  275 &  149.83878 & 2.6750801 & 0.259 & 13.45& 5112  &149.83847 & 2.6750875 & 1608  & 0.259 & 43.08 &  1   \\
  277 &  150.00462 & 2.63275  & 0.677 & 13.45 & 5091 & 150.00452 & 2.6328416&  616  &  0.678 & 42.73 &  1 \\   
  289 &  150.11256 & 2.5560 & 0.501 &  13.57 & 142     & 150.10342 & 2.5504889 & 148   & 0.498 &  42.62 & 0 \\
  292 &  150.03307 & 2.5524 & 0.747 &  13.48 & 398     & 150.02638 & 2.5620575 & 991   & 0.745 &  42.82 & 0 \\
  296 &  149.55516 & 2.0020 & 0.894 &  13.90 & 10732 & 149.56131 & 2.0087938 & 3549 & 0.930$^*$ &  42.85 & 0 \\
  298 &  149.78191 & 2.1390 & 0.354 &  13.59 & 63       & 149.78223 & 2.1387713 & 313   & 0.355 &  42.57 & 0 \\
  298 &  149.78191 & 2.13906  & 0.354    &  13.59 &  392 &  149.79369 & 2.1256437 & 679 &   0.353 & 41.98  &  1   \\
  298 &  149.78191 & 2.13906  & 0.354   &   13.59 &  -99  & 149.77357 & 2.141633  & 1498 &  0.354  &    41.34     &         0   \\
  300 &  149.72893 & 2.2373 & 0.381 &  13.47 &  -99    & 149.74792 & 2.253087   & 2876 & 0.356 &  41.06 & 0 \\         
  303 &  149.99364 & 2.2585399 & 0.660 & 13.53 & 19 &   149.99367 & 2.2585886 & 450  &  0.659 & 43.38 &  1   \\
  322 &  150.2254   & 2.26872&0.677 &  13.51 &  -99    & 150.228     & 2.2698331 & 24     & 0.678 &  42.46 & 0 \\             
  324 &  150.02414 & 2.36050&0.726 &  13.39 &  254    & 150.03079 & 2.358371   & 533   & 0.786 &  42.89 & 0 \\
  333 &  150.0423  & 2.6949  &  0.219   &   13.31 & 5075 & 150.04155 & 2.6945302& 623 &   0.221  &    41.46 & 1 \\  
\enddata
\tablenotetext{a}{Mass defined respect to 200 times the mean density, with $h=$0.72.}
\tablenotetext{c}{= -99 means NOT detected.} 
\tablenotetext{c}{Photometric or spectroscopic redshift.}
\tablenotetext{d}{0: AGN among satellite galaxies; 1: AGN among central galaxies.}
\tablenotetext{*}{Photometric redshift from \citet{Sal11} .}
\end{deluxetable*}

%structured as col (1): galaxy groups ID, col (2,3): galaxy groups $\alpha$ and $\delta$, col (4):
%photometric or spectroscopic redshift, col (5): galaxy group mass estimates, col (6):
%AGN ID in XMM-COSMOS catalog, col (7,8): AGN $\alpha$ and $\delta$, col (9): AGN ID in
%C-COSMOS catalog, col (10): spectroscopic or photometric redshift, col (11):
%rest-frame soft X-ray AGN luminosity, col (12): =0 or =1 for AGN among
%satellite or central galaxies, respectively.

\acknowledgments

%We gratefully acknowledge the contributions of the entire COSMOS collaboration consisting of more than 100 scientists. 
%More information on the COSMOS survey is available at {\url{http://www.astro.caltech.edu/~cosmos}}. 
%%We thank the anonymous referee for carefully reading the manuscript and providing us with constructive remarks.
%V.A., G.H. \& M.S. acknowledges support by the German Deutsche Forschungsgemeinschaft,
%DFG Leibniz Prize (FKZ HA 1850/28-1).
%This work is partially supported by the CONACyT Grant Apoyo 83564.

\clearpage

\end{document}